%% file: template.tex
\documentclass[journal]{vgtc}                     

\usepackage{amsmath}
\usepackage{xspace} 

\usepackage{url}
\usepackage{tikz}
\usepackage[most]{tcolorbox}
\usepackage{ragged2e}   
\usepackage{listings}  
\usepackage{booktabs}
\usepackage{tabularx}
\usepackage{listings}
\usepackage{xcolor}
\usepackage{booktabs}
\usepackage{multirow}
\usepackage{graphicx}

\usepackage{xcolor}

\definecolor{promptbg}{gray}{0.95}
\definecolor{promptframe}{gray}{0.70}

\usepackage{selectp}

\lstdefinelanguage{json}{
    basicstyle=\small\ttfamily,
    columns=fullflexible,
    breaklines=true,
    stringstyle=\color{blue},
    identifierstyle=\color{black},
    keywordstyle=\color{magenta},
    frame=single,
    backgroundcolor=\color{gray!5},
    showstringspaces=false
}

\definecolor{StepCColor}{HTML}{696565}

\newcommand*\roundedbox[1]{\tikz[baseline=(char.base)]{
            \node[shape=rectangle,fill=StepCColor, text=white, draw=StepCColor,   
      line width=1pt,  inner sep= 1.2pt, minimum size=8pt,rounded corners=1pt] (char) {\textbf{#1}}}}

\newcommand{\sysname}{\textit{VizPilot}\xspace}

\onlineid{1723}

\vgtccategory{Research}

\vgtcpapertype{Representations \& Interaction }

\title{VizPilot: Automated Onboarding for SVG-based Composite Visualizations using Multimodal LLMs}

\author{
  \authororcid{Nishaanthini Gnanavel}{0009-0005-6753-1732}
  and
  \authororcid{Yong Wang}{0000-0002-0092-0793}
}
\authorfooter{
\item
Nishaanthini Gnanavel and Yong Wang are with Nanyang Technological University.
E-mail: GN0001NI@e.ntu.edu.sg, yong-wang@ntu.edu.sg.
}
\abstract{%
Composite visualizations integrate multiple visualizations in a meaningful manner and show strong expressiveness in representing complex dataset. They have been widely proposed by visualization developers in different applications.
However, they often impose a high initial cognitive load on novice users, due to their intrinsic composite designs.
Existing visualization onboarding approaches have attempted to mitigate this issue, but they 
are typically platform-dependent, require substantial manual authoring effort, and struggle with the structural complexity of composite visualizations, making them unable to be used for general composite visualizations. In this paper, we present \sysname, a novel automated visualization onboarding approach that reverse-engineers composite visualization structure to enable an interactive onboarding experience directly from raw visualization artifacts. The proposed approach consists of two primary modules: the \textit{Composite Visualization Analyzer} and the \textit{Onboarding Interface}. Leveraging Multimodal Large Language Models (MLLMs), the Analyzer employs a two-stage pipeline: The \textit{Semantic Inference} stage decomposes the visualization into individual visual components and extracts structured knowledge using a five category taxonomy to generate atomic explanation units, and the \textit{Semantic Mapping} stage anchors these explanations to precise SVG elements via hierarchical DOM traversal, enabling reliable highlighting and interactions. We implement VizPilot as a browser extension. To initiate the workflow, a visualization developer provides a brief description and optional interaction source code and \sysname then automatically extracts the SVG and generates bitmap image to drive the multimodal reasoning. Then, the interactive Onboarding Interface will show the structured outputs from the Composite Visualization Analyzer, which supports both \textit{Narrative Scrollytelling}, providing guided, step-by-step explanations with synchronized highlighting, and \textit{Free Exploration} for on-demand inquiry by users. We extensively evaluate \sysname through a comparative analysis of different input modalities to assess the pipeline robustness, a usage scenario to illustrate reduced authoring effort, and a user study to confirm its effect in reducing users' cognitive load.
The overall results demonstrate the effectiveness and usability of \sysname in facilitating an easy onboarding of composite visualizations.
}


\teaser{
  \centering
  \includegraphics[width=\linewidth, alt={}]{figs/Vizpilot_Overview.png}
  \caption{\textbf{\sysname Overview.} (A) \sysname consumes multimodal inputs including SVG code, bitmap images, developer-provided description and optional interaction source code (example chart from Highcharts \cite{highcharts_multi_axes}). (B) The Composite Visualization Analyzer processes these via two stages: (B1) Semantic Inference decomposes the chart to extract knowledge and generate explanation text, producing (B3) intermediate semantic representation. (B2) Semantic Mapping then performs Hierarchical Selector Inference to map these units to relevant SVG elements. (C) The resulting (B4) mapped explanations power a dual-mode Onboarding Interface featuring (C1) Narrative Scrollytelling and (C2) Free Exploration.}
  \label{fig:teaser}
}

\graphicspath{{figs/}{figures/}{pictures/}{images/}{./}} 

\usepackage{tabu}                      
\usepackage{booktabs}                  
\usepackage{lipsum}                    
\usepackage{mwe}                       

\usepackage{mathptmx}    
\usepackage{multirow}

\begin{document}



\input{source/introduction}

\input{source/related_work}
\input{source/methodology}
\input{source/Evaluation}

\input{source/Discussion}

\input{source/conclusion}

\bibliographystyle{abbrv-doi-hyperref}


\bibliography{template}

\clearpage

\input{source/appendix}







\end{document}

%% file: source/introduction.tex
\firstsection{Introduction}
\maketitle


Composite visualizations have become a popular visualization design strategy in the past decades~\cite{javed2012exploring,deng2022revisiting,zhu2024compositingvis}. With the growing size and complexity of datasets, common basic charts like bar charts, pie charts and line charts are overwhelmed and unable to tackle complex analytics tasks in different applications. Accordingly, visualization researchers and developers have proposed various novel composite visualizations by integrating multiple visual representations in a coherent and meaningful manner.
Such composite visualizations can facilitate a comprehensive exploration of different data facets and their relationships, enabling advanced data analysis. A large number of composite visualizations have been presented in the field of data visualization and more are expected to be developed for different application domains in the future~\cite{deng2022revisiting}.


However, despite their expressiveness and popularity, novice users often encounter a significant cognitive barrier when starting to comprehend and use composite visualizations. Common questions, like what each component represents and how interactions between components work, will arise. Multiple visualizations can be combined in different ways such as juxtaposition, overloading, superimposition and nesting~\cite{javed2012exploring}, and different visual encoding channels are used to represent complex and multivariate datasets as well as their relationships. Also, interactions are often enabled for composite visualizations, making it even more challenging for novice users to understand and interact with a composite visualization for the first time. 
%
For example, MatrixWave~\cite{zhao2015matrixwave} (Fig.~\ref{fig:onboardingmodes}) is a composite visualization integrating multiple matrix visual representations to display event sequences, which requires users to simultaneously interpret matrix-based transitions, spatial layouts, and interaction-driven highlighting.
PrettiSmart~\cite{wen2025prettismart} (Fig.~\ref{fig:usagescenario}) is another composite visualization that juxtaposes multiple barcode-based designs to overview smart contract simulation results, and links each barcode-based design to a flow-based design to display transaction details.
When first encountering customized composite visualizations, users must infer component encodings, relationships, and linked interactions, which is cognitively demanding. {\color{black}These challenges are particularly acute for visualization consumers (e.g., researchers, analysts, and domain practitioners) encountering customized composite visualizations without accompanying guidance. They typically rely on manual inspection or consulting the original authors. To better support these consumers, visualization developers need tools to generate onboarding materials.} 



Different visualization onboarding approaches~\cite{stoiber2019visualization,dhanoa2022process} have been proposed to help visualization users easily understand and use an unfamiliar visualization. Specifically, prior research on visualization onboarding has explored the manual creation of interactive tours via authoring~\cite{hoque2025dashguide}, semi-automatic generation of guided tours~\cite{dhanoa2024d}, and a multimodal assistant to answer users' questions~\cite{dhanoa2025hey}.
Despite being effective in specific contexts, these approaches share three key limitations that hinder them from widespread adoption. 
First, these approaches require \textit{substantial authoring effort} to enable visualization onboarding, such as manual specifications~\cite{zhao2024leva}, predefined templates~\cite{dhanoa2024d} and recorded interactions~\cite{hoque2025dashguide}. Second, they are typically \textit{platform-dependent} and depend on proprietary APIs or structured metadata exposed by visualization tools such as Tableau or Power BI~\cite{dhanoa2025hey}~\cite{hoque2025dashguide}. 
Third, they predominantly focus on \textit{basic charts or dashboards} consisting of basic charts.
These three key limitations restrict their applicability to real-world visualization practice and make them unable to facilitate effective onboarding of sophisticated composite visualizations.


Visualization developers often create composite visualizations as customized SVG-based visualizations by using programming libraries like D3.js~\cite{bostock2011d3} and Vega-Lite~\cite{satyanarayan2016vega}.
SVG has become the most common format of composite visualizations. Since SVG is directly accessible in the browser DOM, it provides the foundation for possible platform-agnostic onboarding.
Inspired by this, 
we propose \sysname, an automated visualization onboarding approach implemented as a browser extension that requires no manual authoring effort. \sysname comprises two main modules: the \textit{Composite Visualization Analyzer}, a two-stage pipeline designed to bridge the semantic gap between low-level geometry and high-level intent, and the \textit{Onboarding Interface}, which presents the resulting onboarding steps to users (Fig.~\ref{fig:teaser}). Within the Composite Visualization Analyzer, \textit{Semantic Inference} decomposes the composite visualization, extracts structured knowledge, and generates atomic explanation units, while \textit{Semantic Mapping} automatically maps these units to precise SVG elements. To initiate the process, a developer provides a brief description about the visualization and optional interaction source code and \sysname then extracts the raw SVG code and bitmap image to perform MLLM-based reasoning. The Onboarding interface delivers both \textit{Narrative Scrollytelling} for guided, step-by-step explanations with relevant element highlighting and \textit{Free Exploration} for user-driven interrogation through hover-based inspectors and an AI-driven exploration assistant.

 Given a SVG-based composite visualization, VizPilot automates visualization onboading while still keeping developers' agency. Developers can review and refine generated explanations directly on the interface without altering the underlying code,and export the finalized onboarding which can be seamlessly integrated into their existing project. At runtime on the end-user side, VizPilot dynamically overlays interactive onboarding interface onto existing visualizations, enabling users to interpret composite visualizations without any modifications to the underlying code.

Our main contributions are as follows:
\begin{itemize}
    \item We propose \textit{Composite Visualization Analyzer}, an effective technique to reconstruct composite visualizations from raw artifacts and produce semantically-meaningful explanation units for the corresponding visual elements of a composite visualization, i.e., a two-stage pipeline comprising \textit{Semantic Inference} and \textit{Semantic Mapping}.
    \item We develop \sysname, a browser extension that leverages this Analyzer to drive an \textit{Onboarding Interface}, where it automatically provides synchronized \textit{Narrative Scrollytelling} and context-aware \textit{Free Exploration}, allowing developers to deploy interactive onboarding without any manual authoring effort.
    \item We conduct extensive evaluations to demonstrate the effectiveness and usability of VizPilot in visualization onboarding for both visualization developers and users, which includes a comparative analysis of input modalities on inference accuracy, a usage scenario, a gallery of diverse examples and an in-depth user study with 16 visualization users. We have published our example gallery online: \url{https://vizpilot-gallery.onrender.com}.
\end{itemize}


%% file: source/related_work.tex
\section{Related Work}

This work is related to prior research on visualization onboarding, semantic understanding of visualization, and LLM-based visualization reasoning.
\subsection{Visualization Onboarding}

Visualization onboarding supports users in interpreting and extracting information from visual data representations~\cite{stoiber2019visualization}. Effective onboarding addresses knowledge gaps across domain, data, visual encoding, and interaction through structured storytelling~\cite{stoiber2019visualization, stoiber2022perspectives, stoiber2023design}. Research suggests that users prefer in-situ scrollytelling over static videos, as it reduces cognitive load and the split-attention effect~\cite{stoiber2022comparative}. \sysname adopts these proven principles to bridge those core knowledge gaps.

To operationalize these guidelines, recent work has explored semi-automated onboarding and narrative authoring tools. Dashboard systems such as Hey Dashboard!~\cite{dhanoa2025hey}, DashGuide~\cite{hoque2025dashguide}, and D-Tour~\cite{dhanoa2024d} leverage dashboard metadata to generate guides. Other approaches support slideshow, scene-based, comic-style, scroll-driven, annotated and animated visual narratives~\cite{wang2018narvis,satyanarayan2014authoring,zhao2021chartstory,morth2022scrollyvis,li2023geocamera,wang2024wonderflow,10.1145/3313831.3376443}. LEVA~\cite{zhao2024leva} and the Authoring Tool for Data Journalists~\cite{stoiber2023authoring} instead rely on declarative grammars (e.g., Vega-Lite) to generate template-based explanations.

These approaches remain platform-dependent, authoring intensive, and limited in composite visualization support. \sysname addresses these gaps by operating directly on SVG artifacts and leveraging MLLMs for multimodal semantic analysis, enabling a platform-agnostic pipeline that interprets composite visualizations without manual authoring or access to underlying specifications.

\subsection{Semantic Understanding of Visualization}
Semantic understanding recovers the logic, data mappings, and functional roles of graphical primitives, often using a multi-level taxonomy that decomposes a chart into structural components (e.g., axes, legends), visual encodings (e.g., marks and channels), and analytical insights. Prior work has approached this via heuristic-based structural analysis and, more recently, deep learning and generative models.

To recover these semantics from low-level primitives, several systems leverage the hierarchical structure of Scalable Vector Graphics (SVG). Mystique~\cite{chen2023mystique} deconstructs SVG charts into a structured representation (GREC) for layout reuse, while others use SVG tree structures as proxies for visual similarity~\cite{li2022structure}. Similarly, SeeChart~\cite{alam2023seechart} analyzes D3-based charts to generate accessible audio summaries. However, these rule-based approaches are generally limited to predefined chart types and struggle to generalize to structurally complex composite visualizations that integrate multiple visualizations within a single interface.

Recent work has incorporated large-scale semantic datasets and generative models. VISANATOMY~\cite{chen2025visanatomy} introduced an SVG corpus and taxonomy for chart semantics, while subsequent work leveraged GNNs and LLMs for video narration~\cite{ying2024reviving}, semantic extraction from infographic timelines~\cite{zhu2019towards}, and reverse-engineering SVG visualizations into semantic representations~\cite{xie2025datawink}.

A key gap remains as existing methods focus on isolated charts, missing the coordinated semantics and cross-component relationships of composite visualizations, and lack deterministic grounding for interactive highlighting. We address this with a multimodal pipeline that fuses hierarchical SVG structures and bitmap images to reconstruct component-level organization, mapping structured knowledge directly to SVG elements for precise, interactive onboarding.


\subsection{LLM-Based Visualization Reasoning}

Recent research has investigated the capacity of Large Language Models (LLMs) and Multimodal LLMs (MLLMs) to interpret and reason about visual data. Instruction-tuned systems like ChartLlama~\cite{han2023chartllama} excel at chart-based QA and summarization~\cite{do2023llms,zeng2024advancing}, while systematic benchmarks across GPT-4, Claude, and Gemini on VLAT~\cite{7539634} and CALVI~\cite{10.1145/3544548.3581406} show that MLLMs can interpret common encodings and identify high-level trends~\cite{khan2025evaluating, pandey2025benchmarking, bendeck2024empirical,hong2025llms}.

Despite these advances, MLLMs often struggle with precise numerical extraction, subtle encoding distinctions, and identifying misleading visualizations~\cite{khan2025evaluating, pandey2025benchmarking}. While structured prompting strategies like Charts-of-Thought~\cite{das2025charts} partially mitigate these issues through intermediate reasoning, models still face difficulties with precise quantitative reasoning and cross-view coordination when processing raw SVG representations~\cite{xu2024exploring}. These limitations highlight the fragility of current multimodal reasoning approaches without explicit structural support.

Existing methods focus mainly on single-chart understanding via static benchmarks, producing free-form text outputs with no structural alignment to graphical elements. \sysname instead frames visualization interpretation as a constrained multimodal pipeline, decomposing composite visualizations and aligning semantic roles directly with low-level SVG elements to support interactive onboarding.

%% file: source/methodology.tex

\section{\sysname: Overview}

Our goal is to automatically generate onboarding guidance for composite visualizations without requiring manual annotations or access to the visualization specifications. Rather than relying on declarative grammars or platform-specific APIs, our approach reconstructs visualization semantics directly from raw visualization artifacts.

Figure \ref{fig:teaser} illustrates the overall workflow of \sysname. Given a web-rendered visualization, \sysname extracts the SVG code, generates the bitmap image. To further refine the semantic inference, visualization developers can provide a brief high-level description about the visualization and optionally upload the interaction source code. These multimodal inputs are processed through a pipeline that reconstructs the visualization's semantic structure and generates mapped explanation units for onboarding.

The methodology consists of two main components:

\begin{itemize}
    \item \textbf{Composite Visualization Analyzer:} A two-stage pipeline consisting of \textit{Semantic Inference} to decompose composite visualizations and extract structured knowledge, and \textit{Semantic Mapping} to map these inferred explanations directly to SVG elements for precise interactive highlighting.
    \item \textbf{Onboarding Interface:} A dual-mode onboarding that transforms this structured representation into an actionable user experience, supporting both guided \textit{Narrative Scrollytelling} and context-aware \textit{Free Exploration}.
\end{itemize}

Together, these components enable automated onboarding generation directly for SVG-based composite visualizations, while maintaining precise alignment between textual explanations and the underlying visual elements.

\section{Composite Visualization Analyzer}
We formalize Composite Visualization Analyzer 
as a reasoning process that reconstructs the semantic structure of composite visualizations from raw artifacts to support onboarding. Prior work such as DataWink \cite{xie2025datawink} shows that multimodal models can derive semantically enriched representations from SVG-based visualizations, but primarily focuses on reuse and template generation rather than supporting user onboarding. However, interpreting web-based composite visualizations directly from raw artifacts remain challenging: SVG representations encode low-level graphical primitives without exposing high-level constructs such as views or analytical relationships, existing approaches often depend on platform-specific APIs which limits generalizability, interactive behaviours (e.g., filtering or brushing) are defined in scripts and are not evident from static artifacts, and generative methods often struggle to align explanations with visual elements.

To address these challenges, we adopt a knowledge-driven approach that decomposes a visualization into atomic explanation units. Rather than generating free-form descriptions directly, Composite Visualization Analyzer first constructs an intermediate semantic representation capturing visual components, knowledge types, and explanation order, which is then mapped to SVG elements for interaction and highlighting. Generating accurate onboarding cannot be achieved in a single stage because semantic interpretation and SVG element alignment require fundamentally different reasoning. We therefore decompose the analyzer into two stages (Fig.~\ref{fig:teaser}\roundedbox{B}): \textbf{Semantic Inference}, which reconstructs high-level concepts such as components, encodings, and interactions from multimodal inputs to produce structured explanation units, and \textbf{Semantic Mapping}, which maps these units to specific visual elements to enable precise highlighting and interaction during onboarding.

To support this two-stage reasoning process, the pipeline operates over multimodal inputs, including SVG code for hierarchical structure and element localization, rendered bitmap images to capture spatial layout and context missing from SVG code, author-provided descriptions to guide semantic interpretation and narrative focus, and interaction source code (e.g. D3.js) to help disambiguate interactions. Together, these inputs enable our approach to progressively transform raw visualization artifacts into mapped, structured explanation units for onboarding.

\subsection{Semantic Inference}

The semantic inference stage performs the core reasoning required to reconstruct the semantic structure of the visualization from the collected inputs. Instead of directly producing the free-form textual descriptions, this stage derives a structured intermediate representation that organizes the visualization into components and associated explanation units that can later be mapped to specific visual elements. To reduce hallucinations, reasoning follows a constrained multi-step prompting strategy in which each step's output bounds the next. This stage is decomposed into three steps to reflect the hierarchical nature of visualization understanding: it must first identify what constitutes meaningful units in the visualization, then determine what each unit represents, and finally express this information in a form suitable for onboarding. Accordingly, it consists of (1) \textbf{Visual Component Decomposition}, which segments the visualization into logical visual components using spatial and structural cues; (2) \textbf{Knowledge Extraction}, which derives structured semantic information for each visual component based on a predefined taxonomy; and (3) \textbf{Semantic-to-Text Generation}, which converts this structured knowledge into concise, atomic explanation units (Fig.~\ref{fig:teaser}\roundedbox{B1}). These three steps are executed within a single structured prompt to enable joint reasoning.

    
    



\paragraph{\textbf{Visual Component Decomposition.}}
Composite visualizations often contain multiple integrated visualizations, each with distinct semantic and functional roles, such as charts, legends, axes, or overview panels. SVG code provides low-level graphical primitives and hierarchical grouping, but it does not explicitly encode high-level component boundaries or inter-component relationships. To address this, the Component Decomposition step identifies and organizes these visual structures into meaningful \textbf{\textit{visual components}} by combining information from the rendered bitmap image and the SVG structure to perform decomposition. Spatial layout cues from the image, such as the relative positions, adjacency and alignment, are integrated with the SVG's hierarchical groupings to cluster low-level elements into coherent visual components. 

{\color{black}
We define a visual component as a coherent visual structure that serves as the fundamental unit of decomposition for subsequent knowledge extraction and explanation generation. A visual structure qualifies as a visual component only if it independently supports the extraction of at least one distinct form of \textit{visualization knowledge}~\cite{stoiber2019visualization} that will be introduced in Table~\ref{tab:knowledge_types} and the subsequent paragraph \textit{Knowledge Extraction}. 
For example, an individual axis tick is not considered a visual component because it does not contribute independently extractable knowledge beyond that provided by the axis. 
Also, we do not consider basic visual marks (e.g., a bar in a bar chart) as visual components, because they can share redundant, near-identical visualization knowledge. For example, all the bars in a bar chart will share the same visual encoding knowledge.
It is possible that one visual component can contain multiple other visual components. 
For example, in the PrettiSmart visualization (Fig.~\ref{fig:usagescenario}), the simulation panel (Fig.~\ref{fig:usagescenario}\roundedbox{A1}) constitutes a higher-level visual component that contains multiple simulation thumbnails (Fig.~\ref{fig:usagescenario}\roundedbox{A2}), each of which can also be treated as an individual visual component.
Visual components are categorized into two types. \textit{Data-driven visual components} are coherent visual structures that represent analytical representations of the underlying data that require an independent explanation. Separate data-driven components sharing the same coordinate system are treated independently when they encode different information that requires distinct explanations. \textit{Auxiliary visual components} such as legends, axes, and overview panels provide contextual or reference information that supports interpretation and independently contribute 
extractable knowledge. For instance, the state legend of the PrettiSmart visualization (Fig.~\ref{fig:usagescenario}\roundedbox{A3}) can be treated as an auxiliary visual component.
}

%
This decomposition enables \sysname to reason about each visual component independently, facilitating targeted knowledge extraction and the generation of atomic explanation units. By mapping subsequent analysis in well-defined components, \sysname reduces ambiguity and supports multi-component co-ordination. Further justification for the visual component decomposition strategy is provided in Appendix~\ref{subsec:component-decomposition}.

\paragraph{\textbf{Knowledge Extraction.}}
Without an explicit visualization grammar, it is difficult to programmatically determine what data is shown, how it is encoded, and what analytical meaning it conveys. To prevent hallucinations or generic descriptions, this step enforces a strict five-category knowledge taxonomy derived from visualization onboarding literature \cite{stoiber2019visualization}: Structural, Visual Encoding, Data, Analytical and Interaction knowledge. 
As summarized in Table \ref{tab:knowledge_types}, the model evaluates each decomposed visual component and extracts relevant entities, classifying them into the predefined knowledge categories.

\begin{table}[tb]
\renewcommand{\arraystretch}{1.4} %
  \caption{%
    Types of knowledge and their inference objectives.%
  }
  \label{tab:knowledge_types}
  \scriptsize
  \centering
  \begin{tabu}{p{2cm} p{5.6cm}}
    \toprule
    \textbf{Knowledge Type} & \textbf{Inference Objective} \\
    \midrule
    Structural Knowledge & Captures the overall layout, hierarchical relationships, and shared structural components such as axes or legends. \\

    Data Knowledge & Extracts the underlying domain semantics, encompassing the specific variables, units, and categorical contexts represented within the visualization. \\

    Visual Encoding Knowledge & Explicates the specifc mappings between data attributes and graphical primitives (e.g., position, colour, size) within the decomposed components. \\

    Analytical Knowledge & Synthesizes data trends, comparisons and inter-component relationships to articulate how multiple components coordinate to generate insights. \\

    Interaction Knowledge & Deduces the supported interactions (e.g., hover, click, brush, zoom), explaining the analytical purpose of these user actions. \\
    \bottomrule
  \end{tabu}
\end{table}

Furthermore, interactive behaviors (e.g., brushing and linking) are invisible in the static SVG files, which makes it impossible to infer the interaction knowledge of a composite visualization from the SVG files and unable to provide onboarding explanations for interactions.  To address this, we selectively incorporate interaction source code into the context only to infer the existence and intent of interactions. MLLM is instructed to abstract away the implementation details to ensure that the interaction knowledge is expressed solely from the user's perspective.

This process produces a set of structured visual component-level knowledge units, where each unit describes the visual component type, associated data, visual encodings, and inferred interactions. These units produces a semantically rich representation that serves as input for the subsequent Semantic-to-Text Generation step, ensuring explanations are accurate, user-centric, and aligned with the visualization.


\paragraph{\textbf{Semantic-to-Text Generation.}}
Once structured knowledge has been extracted for each visual component, this step converts this information into concise explanation units suitable for onboarding. This step addresses the challenge that standard MLLM outputs tend to be overly long, mix multiple knowledge categories, and lack suitability for stepwise, progressive disclosure in composite visualizations.

To mitigate this issue, this step generates \emph{atomic explanation units}, each describing a single concept associated with one view and one knowledge category. By constraining each unit to exactly one sentence and one knowledge type, the pipeline ensures clarity and cognitive accessibility for first-time users. This rendering process also enforces a hierarchical presentation order to guide users logically from understanding the layout and components of the visualization to interpreting data, encodings, analytical insights, and interactions, supporting a structured learning experience.

The resulting output is an ordered sequence of atomic explanation units with each corresponding to a visual component, and a concise explanation text. This intermediate semantic representation serves as the input for the semantic mapping stage, ensuring that textual explanations can be reliably mapped to visual elements for interactive onboarding.

\subsection{Semantic Mapping}

\begin{figure}[tb]
    \centering
    \includegraphics[width=8cm]{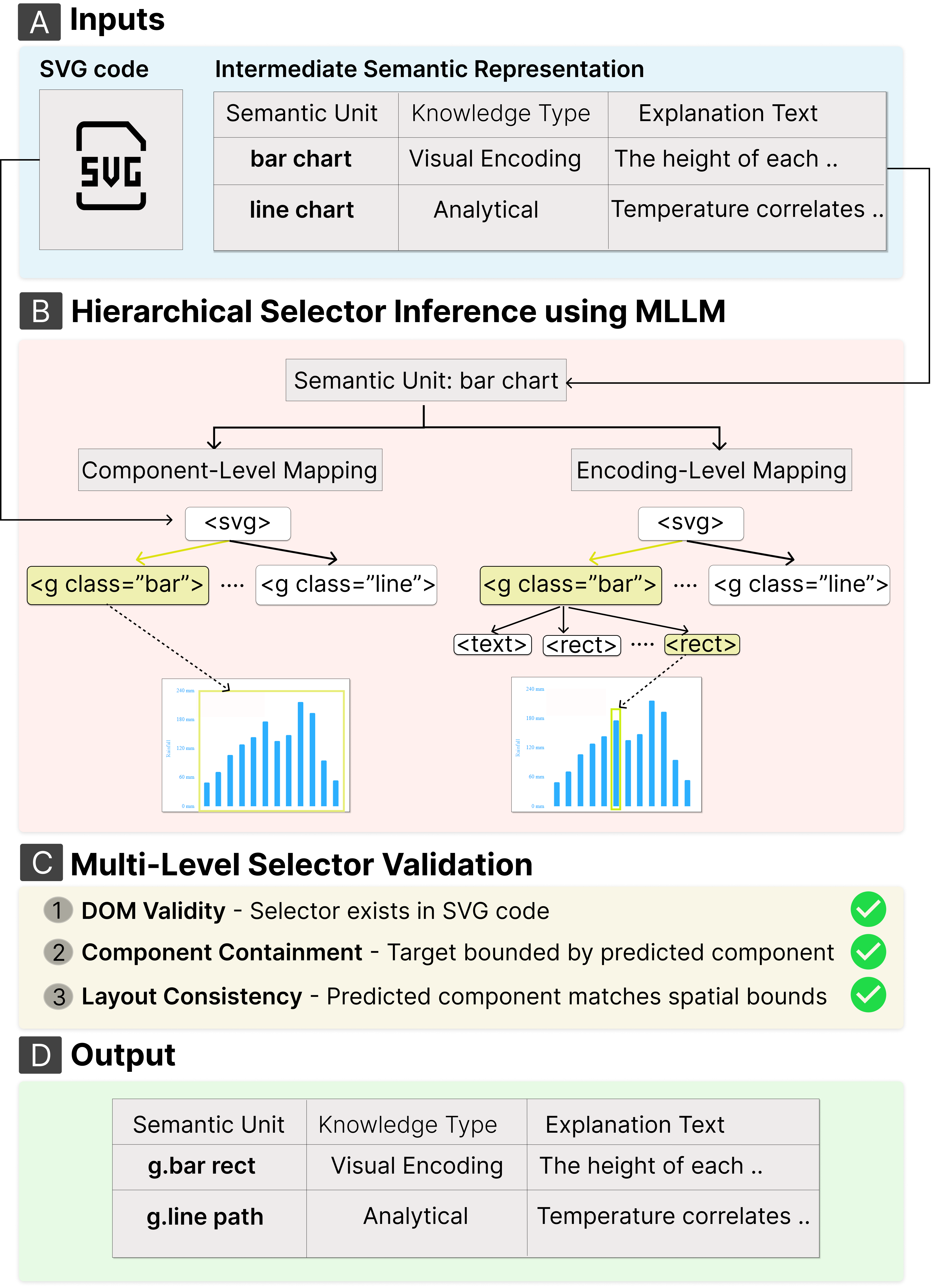} 
    \caption{\textbf{The Semantic Mapping Stage.} \sysname maps each semantic unit from the (A) Intermediate Semantic Representation into precise SVG elements. (B) An MLLM proposes candidate DOM selectors through a hierarchical selector inference for both component-level and encoding-level targets, which are then (C) verified against a multi-level selector validation. The resulting (D) mapped explanation units enable synchronized visual highlighting for the onboarding interface.}
    \label{fig:semanticmapping}
\end{figure}

While Semantic Inference produces structured explanation units, these units remain abstract until they are visually anchored to the corresponding elements in the visualization. It is non-trivial to build the mapping between structured explanations and visualization elements, as raw SVG code often contains generic groupings, inconsistent naming, and deeply nested structures, making the direct alignment difficult.

We address this with \textit{Semantic Mapping}, a mapping stage that links explanation units to SVG elements using the intermediate representation and SVG code (Fig.~\ref{fig:semanticmapping}\roundedbox{A}). Rather than relying on zero-shot generation, we designed a specialized prompting strategy that explicitly instructs the MLLM to act as an expert DOM parser. Our key contribution is a \emph{Hierarchical Selector Inference} mechanism where instead of directly outputting element IDs, the MLLM is prompted to construct selector paths from parent containers to target nodes, enabling deterministic and interpretable grounding. The sematic mapping here operates at two levels of granularity:
\begin{itemize}
    \item \textbf{Component-Level Mapping:} The MLLM first identifies the macro-structure, mapping structural descriptions to the correct bounding container (e.g., the top-level \texttt{<g>} defining a specific chart panel).
    \item \textbf{Encoding-Level Mapping:} Constrained within the resolved container, the MLLM then links fine-grained explanation units to specific visual elements (e.g., data marks, legends, or axis ticks), supporting accurate highlighting and interaction.
\end{itemize}

During this process, the MLLM resolves ambiguities in mapping by leveraging contextual information such as component hierarchy, relative positioning, and layout structure to disambiguate candidate elements. 
For the nodes without unique identifiers, our approach generates hierarchical selectors to enable deterministic targeting, and favors selectors that uniquely identify elements to avoid ambiguous mappings.

Since MLLMs are probabilistic, relying solely on their output for UI targeting is unsafe. All inferred selectors are therefore passed through a programmatic multi-level verification step (Fig.\ref{fig:semanticmapping}\roundedbox{C}). First, selectors are resolved against the SVG DOM to ensure \textbf{DOM Validity}. Next, retrieved elements are checked for \textbf{Component Containment}, verifying consistency with the predicted component decomposition. Finally, \textbf{Layout Consistency} is enforced by comparing the spatial extent of mapped elements with the bounding regions of candidate containers derived from the SVG layout. Selectors that fail any of these checks are rejected and regenerated.

By combining MLLM-based hierarchical inference with programmatic validation, this stage produces precise and verifiable mappings between abstract explanation units and SVG elements (Fig.\ref{fig:semanticmapping}\roundedbox{D}), enabling interactive onboarding features such as stepwise highlighting and hover-based exploration.

\section{Onboarding Interface}

\begin{figure}[tb]
    \centering    \includegraphics[width=\columnwidth]{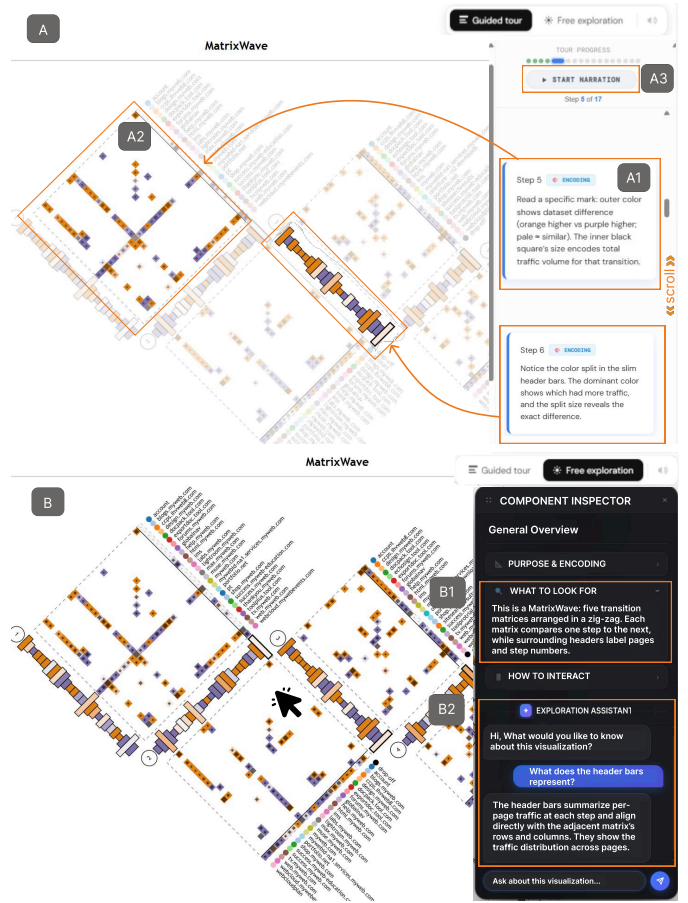} 
\caption{\textbf{Onboarding Interface in MatrixWave}~\cite{zhao2015matrixwave}. (A) Narrative Scrollytelling provides a structured, linear guide where each narrative step (A1) automatically highlights the relevant visual element (A2), accompanied by text-to-speech narration (A3). (B) Free Exploration supports user-driven inquiry where hovering or selecting a component triggers an immediate semantic overview (B1), which can be further be investigated via the conversational exploration assistant (B2).}
\label{fig:onboardingmodes}
\end{figure}


The Onboarding Interface operationalizes the structured output of the Composite Visualization Analyzer, transforming mapped explanation units into an interactive onboarding experience. While the Analyzer resolves what the visualization represents, the interface must determine how this knowledge is delivered to support effective onboarding. This introduces a tension between providing scaffolded guidance to reduce cognitive load and preserving user agency for flexible exploration, as purely guided or unguided approaches are each insufficient.

To address this, we adopt a dual-mode design that separates guided explanation from exploratory interaction. The \textbf{Narrative Scrollytelling Mode} presents explanation units sequentially with synchronized highlighting to build an initial mental model, while the \textbf{Free Exploration Mode} enables independent interaction supported by an Exploration Assistant that provides contextual, on-demand guidance. This design supports both structured onboarding and flexible analysis while maintaining alignment between explanations and visual elements.

\paragraph{\textbf{Narrative Scrollytelling Mode.}}
The Narrative Scrollytelling mode (Fig. \ref{fig:onboardingmodes} \roundedbox{A}) serves as an initial guided introduction to composite visualization. Its primary cognitive objective is to reduce the feeling of overwhelm by establishing the global structure and giving the user a start on where and how to start reading the composite visualization. This mode provides a progressive walkthrough using atomic explanation units presented as scroll-triggered steps where each step focuses on a single knowledge type (Fig.\ref{fig:onboardingmodes}\roundedbox{A1}). To ensure high-fidelity grounding, the interface utilizes the mapped SVG selectors derived during the Semantic Mapping stage to dynamically highlight target elements while dimming unrelated visual components (Fig.\ref{fig:onboardingmodes}\roundedbox{A2}). This precise spatial anchoring ensures that every narrative explanation is visually verified in the visualization. Onboarding steps are ordered hierarchically, where it first introduces visual component roles and data encodings before surfacing analytical trends or complex interactions. This automated scrollytelling delivers a low-friction, scaffolded entry into composite visualizations.


\paragraph{\textbf{Free Exploration Mode.}}
While Narrative Scrollytelling provides a structured introduction, users inevitably require non-linear inquiries to test hypotheses, revisit specific components, or seek deeper semantic clarification. To support this, this Free Exploration mode (Fig.\ref{fig:onboardingmodes}\roundedbox{B}) allows users to interrogate the composite visualization autonomously. Hovering over or selecting a visual component retrieves its mapped semantic entity from the pre-computed representation, displaying a concise overview (Fig.\ref{fig:onboardingmodes}\roundedbox{B1}). To support open-ended questions, a context-constrained chatbot (Fig.\ref{fig:onboardingmodes}\roundedbox{B2}) is initialized with the component’s role, knowledge type, interaction semantics, and rendered image. This multimodal framing anchors responses to pre-inferred semantics, preventing the model from hallucinating or reinterpreting the visualization.

\section{Implementation}
{\color{black}We implemented \sysname as a browser extension, which operates on web-based SVG visualizations that can be accessed through the browser DOM. The browser extension automatically extracts the underlying SVG code and generates the bitmap rendering of the visualization to capture spatial layout context. The developer of a composite visualization is also allowed to provide optional interaction source code, which will be further analyzed by \sysname to improve the accuracy of interaction extraction. While interaction source code is beneficial for recovering interaction semantics, \sysname can still generate onboarding content from the SVG and bitmap inputs alone when no interaction source code is provided.} This deployment model allows \sysname to analyze the rendered composite visualizations in-situ, without requiring any modifications to the source code. To support practical application and workflow separation, our approach distinguishes between onboarding creation and usage through two workflows: the \textit{Authoring Workflow} and the \textit{Consumption Workflow}.

For the \textbf{Authoring Workflow}, visualization developers provide a brief description and optionally the interaction source code. \sysname automatically extracts SVG code, generates bitmap image and leverages GPT-5 to generate the onboarding structure. Developers can review and refine the explanation units before exporting as a decoupled package, which consists of 1) a structured data file in json format with narrative steps and SVG selectors,  and 2) a lightweight JavaScript file for the UI elements. A detailed usage scenario illustrating the authoring workflow and onboarding deployment is provided in Section~\ref{subsec:usage_scenario}. For the \textbf{Consumption Workflow}, the visualization is presented in a read-only setting with onboarding activated in the interface. Visualization users can either follow guided scrollytelling or explore the visualization freely. Optional Text-to-Speech (TTS) narration  further enhances the accessibility of the visual explanations generated by \sysname.

This design enables onboarding for composite visualizations without modifying their original implementation, while supporting a clear separation between authoring and consumption through a reusable, decoupled deployment model.

%% file: source/Evaluation.tex
\section{Evaluation}

We evaluate \sysname through an input modality analysis to assess pipeline robustness, a usage scenario and gallery to demonstrate reduced authoring effort and generalizability, and a user study assessing comprehension and cognitive load.

\subsection{Comparative Analysis of Input Modalities}

We conduct a comparative analysis to isolate the specific contributions of SVG code, bitmap image, and interaction source code to the onboarding pipeline. {\color{black} To validate this, we manually inspect the outputs across four representative visualizations spanning distinct composite design patterns: PrettiSmart, RuleMatrix~\cite{rulematrix}, PonziLens~\cite{10794804}, and Shanghai Index~\cite{echarts_shanghai_index}. Table~\ref{tab:ablation_study} summarizes the observed capabilities and limitations. To illustrate these trade-offs concretely, the detailed inference results for PrettiSmart (PrettiSmart overviews smart contract simulation results via barcode-based summaries, each linked to a flow-based diagram of transaction details) (Fig.~\ref{fig:usagescenario}) are presented below. The detailed results for the remaining
three visualizations are provided in Appendix \ref{subsec:modality-results}.} 

\textbf{SVG code only:} While the raw SVG code captures structural boundaries and provides precise selectors, it lacks visual context, reducing output to geometric description. For example, without visual rendering, the model identifies the tan-colored blocks in the visualization as ``tan blocks'' rather than as ``Payable Functions,'' since the color to category mapping is only conveyed visually.

\textbf{Image only:} This supports high-level semantic reasoning and pattern recognition but lacks structural grounding. The descriptions rely on ambiguous visual anchors like ``wide pink band''  that cannot be mapped to the DOM elements, making the onboarding output unactionable for scrollytelling.

\textbf{SVG code + Image:} Combining these modalities enables coordinated layout identification and valid selector linking. Without interaction logic, \sysname infers interaction behaviors from typical UI appearance rather than verified code. (e.g., assuming hover behaviors).

\textbf{SVG code + Image + Interaction source code:}  Incorporating interaction source code enables precise action-consequence reasoning. By interpreting event handlers directly, \sysname generates guidance grounded in implementation logic, such as identifying that clicking a simulation thumbnail triggers the detailed Functions and Contract/Address view, shifting onboarding from static description to actionable assistance.

In summary, each modality is essential, where removing images sacrifices semantics, removing SVG negatively affects the DOM mapping accuracy, and removing interaction source code reduces interactions to guesswork.


\begin{table}[tb]
  \renewcommand{\arraystretch}{1.3}
  \setlength{\tabcolsep}{4pt} 
  \caption{Capabilities and limitations of different input configurations.}
  \label{tab:ablation_study}
  \scriptsize
  \centering
  \begin{tabular}{>{\raggedright\arraybackslash}p{2cm} >{\raggedright\arraybackslash}p{3.4cm} >{\raggedright\arraybackslash}p{3.4cm}}
    \toprule
    \textbf{Input} & \textbf{Capabilities} & \textbf{Limitations} \\
    \midrule
    \textbf{SVG code Only} & 
    Isolates individual components and extracts precise DOM selectors. & 
    Fails to infer visual hierarchy or spatial relationships; focuses on low-level geometry. \\
    \addlinespace
    
    \textbf{Image Only} & 
    Demonstrates visual reasoning and identifies layout patterns. & 
    Lacks syntactic precision; explanations remain ungrounded and unactionable. \\
    \addlinespace
    
    \textbf{SVG code + Image} & 
    Recognizes tiered layouts and grounds high-level explanations. & 
    Does not fully verify interactive behavior.\\
    \addlinespace
    
    \textbf{SVG code+ Image + Interaction source code} & 
    Verifies event handlers and deduces action-consequence. & 
    Minimal; Primarily dependent on input completeness and code availability. \\
    \bottomrule
  \end{tabular}
\end{table}


\subsection{Usage Scenario: Authoring Workflow for a Composite Visualization}
\label{subsec:usage_scenario}

\begin{figure*}[t]
\centering  
\includegraphics[width=\textwidth]{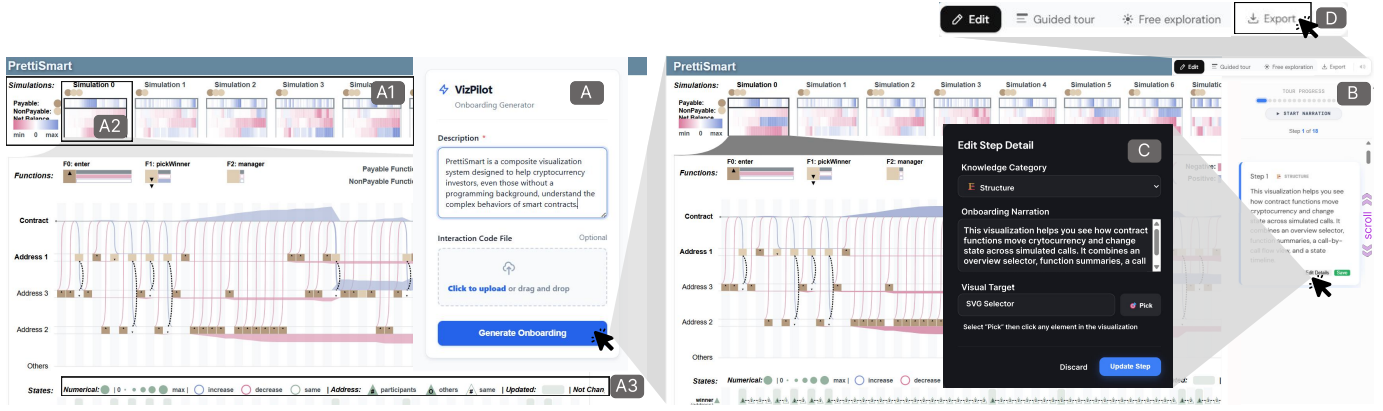}
\caption{\textbf{Usage Scenario: Authoring Workflow for PrettiSmart Composite Visualization}~\cite{wen2025prettismart}. (A) The developer, Alex, provides the interaction source code and a brief description of the visualization. (B) The system automatically generates a mapped onboarding sequence, which Alex reviews for accuracy. (C) Alex refines a specific explanation unit via the edit option. (D) Finally, Alex exports the onboarding content to deploy the onboarding on the visualization.}
\label{fig:usagescenario}
\end{figure*}
To illustrate the authoring workflow, we present a scenario involving Alex, a visualization developer tasked with creating an automated onboarding experience for a complex composite visualization, PrettiSmart (Fig.\ref{fig:usagescenario}). While the visualization contains coordinated charts with rich analytical insights, its intricate structure and hidden interactions present a significant learning curve for first-time users.

\textbf{Automated Generation and Mapping:} Instead of manually coding step-by-step explanations and UI bindings, Alex simply initiates the \sysname browser extension. By providing a brief high-level description and the interaction source code, Alex triggers the Composite Visualization Analyzer. \sysname then automatically extracts the SVG code and generates the bitmap image to perform its two-stage pipeline: it first generates atomic explanation units via Semantic Inference, and then maps those units to precise SVG elements via Semantic Mapping.

\textbf{Review and Refinement:} Using the generated output, \sysname produces onboarding content in narrative scrollytelling and free exploration modes. As Alex reviews the generated onboarding sequence, he observes that each explanation step highlights the relevant visual elements directly within the visualization (Fig.\ref{fig:usagescenario}\roundedbox{B}). Alex then refines one explanation step using the edit option in the interface (Fig.\ref{fig:usagescenario}\roundedbox{C}) without manually adjusting the underlying UI bindings.

\textbf{Decoupled Deployment:} After finalizing the sequence, Alex exports the onboarding specifications. \sysname separates content from rendering logic via a decoupled architecture: a structured json file stores narrative steps and SVG selectors, and a runtime script file handles the UI elements. By including these files in his project, the runtime securely overlays the interactive onboarding onto the visualization, without requiring any modifications to the original visualization code.

\subsection{Gallery}

\begin{figure}[tb]
    \centering    \includegraphics[width=\columnwidth]{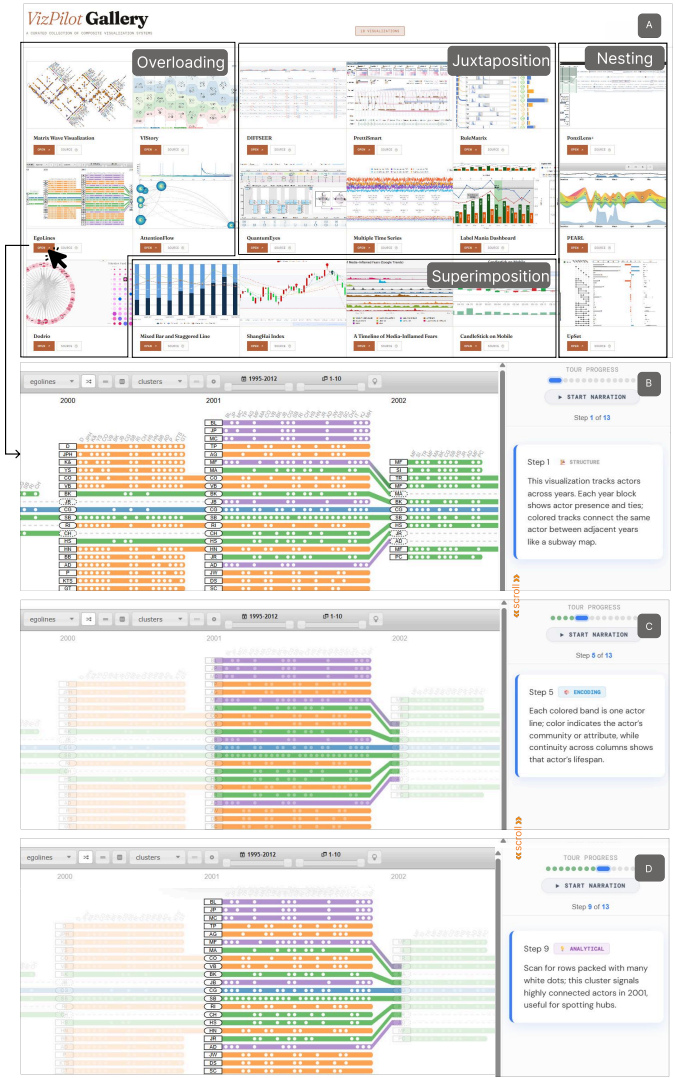} 
\caption{\textbf{Gallery} (A) The \sysname gallery demonstrates generalizability across diverse composite visualizations. A representative onboarding sequence illustrates our narrative scrollytelling mechanism: as the user progresses through (B--D) ordered explanation units, \sysname dynamically highlights the focal component while dimming irrelevant views to mitigate visual clutter and reduce cognitive load.}
\label{fig:gallery}
\end{figure}

{\color{black}To further demonstrate the usefulness of \sysname, we apply it to a diverse set of composite visualization examples\footnote{\url{https://vizpilot-gallery.onrender.com}}. We curated 18 visualizations encompassing all composite design patterns identified in the prior work~\cite{javed2012exploring}: Juxtaposition (e.g., RuleMatrix~\cite{rulematrix}), Superimposition (e.g., ShangHai Index~\cite{echarts_shanghai_index}), Nesting (e.g., PEARL~\cite{pearl}), and Overloading (e.g., VIStory~\cite{vistory}) (Fig.~\ref{fig:gallery}\roundedbox{A}). To quantitatively assess the robustness across this diverse gallery, we evaluated all 18 visualizations across three dimensions: component decomposition accuracy (Precision, Recall, F1), semantic mapping accuracy (exact mapping and partially-correct mapping), and explanation coherence (type alignment, component alignment, atomicity). Evaluation was conducted using an LLM-as-judge protocol~\cite{GU2026101253} with three judge models, GPT-5.5, Gemini 3.1 Pro Preview, and Claude 3.5 Sonnet. Each of the model received the visualization bitmap, SVG code, and full pipeline output and assessed them against a structured rubric (Appendix~\ref{subsec:quantitative-gallery}). Each judge reports a confidence score in [0,1] per criterion and the cases with a confidence score lower than 0.7 in any of the criteria were escalated to manual review (5 of 18) by the authors. Also, we reported the end-to-end (E2E) success rate that requires correct decomposition, mapping, and coherence jointly per ground-truth component. Table~\ref{tab:gallery_eval} reports the results aggregated by each composite design pattern. VizPilot achieves a strong performance consistently across all four patterns. The detailed results for each composite visualization can be found in Appendix~\ref{subsec:quantitative-gallery}.}

As a representative case, we illustrate the onboarding for EgoLines~\cite{zhao2016egocentric}, a composite visualization that integrates continuous actor lines with discrete adjacency blocks for egocentric dynamic network analysis (Fig.~\ref{fig:gallery}\roundedbox{B}--\roundedbox{D}). \sysname automatically generates a 13-step scrollytelling sequence: an initial structural step establishes the hybrid layout and temporal axis (Fig.~\ref{fig:gallery}\roundedbox{B}); the encoding steps selectively emphasize specific components such as actor lines while dimming surrounding context to mitigate visual clutter (Fig.~\ref{fig:gallery}\roundedbox{C}); and the analytical steps highlight complex visual patterns and suggest exploration strategies (Fig.~\ref{fig:gallery}\roundedbox{D}). A visualization user can either follow the guided tour using the Narrative Scrollytelling mode, or switch to Free Exploration mode for on-demand inquiries.

\begin{table}[t]
  \centering
  \caption{Pipeline robustness evaluation aggregated by composite design pattern across 18 gallery visualizations. Decomposition (Precision, Recall, F1); Semantic Mapping (Exact, Partial accuracy); Explanation Coherence (Type alignment, Component alignment, Atomicity); E2E: End-to-End Success Rate.}
  \label{tab:gallery_eval}
  \resizebox{\columnwidth}{!}{%
  \begin{tabular}{l ccc cc ccc c}
    \toprule
    \multirow{2}{*}{\textbf{Composite Pattern}} 
      & \multicolumn{3}{c}{\textbf{Decomposition}} 
      & \multicolumn{2}{c}{\textbf{Semantic Mapping}} 
      & \multicolumn{3}{c}{\textbf{Coherence}} 
      & \textbf{E2E} \\
    \cmidrule(lr){2-4} \cmidrule(lr){5-6} \cmidrule(lr){7-9} \cmidrule(lr){10-10}
    & Prec. & Rec. & F1 & Exact & Partial & Type & Comp & Atom & \\
    \midrule
    Juxtaposition   & 0.95 & 0.96 & 0.96 & 0.93 & 0.98 & 0.96 & 0.99 & 0.94 & 0.93 \\
    Overloading     & 0.96 & 0.88 & 0.92 & 0.89 & 0.95 & 0.99 & 0.97 & 0.86 & 0.83 \\
    Nesting         & 0.96 & 0.90 & 0.93 & 0.92 & 0.98 & 0.96 & 0.96 & 0.89 & 0.91 \\
    Superimposition & 1.00 & 0.94 & 0.97 & 0.92 & 0.96 & 0.95 & 1.00 & 0.94 & 0.94 \\
    \bottomrule
  \end{tabular}%
  }
\end{table}


\subsection{User Study}

To extensively evaluate the effectiveness of \sysname from an end-user perspective, we conducted a controlled user study. Specifically, we aimed to assess whether \sysname improves visualization comprehension and reduces cognitive load when compared to a text-based baseline.

\textbf{Participants.} We recruited 16 participants (5 females, 11 males) aged 20 to 32 years (M = 25.94, SD = 3.91) from analytical disciplines including Computer Science, Data Science, and Mathematics. Participants reported frequent engagement with data visualizations, but only moderate prior familiarity with composite visualizations (7-point Likert scale: M=4.75, SD=1.29). Each one-on-one session was conducted in-person or via Zoom, with participants compensated 10 USD/hour. This study was approved by our institutional IRB, and informed consent was obtained from all participants prior to the experiment. Participants are anonymized as P1–P16 in subsequent analyses.

\textbf{Design and Tasks.} We adopted a within-subjects design with Latin Square counterbalancing to the mitigate ordering effects. Each participant answered four analytical questions on two composite visualizations, MatrixWave (Fig.~\ref{fig:onboardingmodes}) and PrettiSmart (Fig.~\ref{fig:usagescenario}), requiring understanding of structural hierarchy, visual encodings, data trends, and interactions. {\color{black}We selected MatrixWave and PrettiSmart because they represent distinct composite visualization structures and onboarding challenges. MatrixWave requires users to integrate information across coordinated matrix components, whereas PrettiSmart combines heterogeneous visual encodings and interactions across linked panels.} The Baseline condition provided static text descriptions extracted from the corresponding papers for each composite visualization. The Experimental condition provided dual-mode explanations for each composite visualization.

\textbf{Procedure.}  Each session comprised three phrases and lasted about 60 minutes. During the initial 10-minute briefing, participants completed consent and demographic forms, and performed a practice walkthrough of both conditions. In the 35-minute task execution phase, participants explored two composite visualizations under alternating conditions, answering four analytical questions and completing a post-condition Likert questionnaire after each condition. Finally, the session concluded with a 15-minute semi-structured interview to capture qualitative insights regarding participants’ sensemaking strategies and overall feedback.

\textbf{Measurements and Analysis.} To evaluate the effectiveness of \sysname, we measured task accuracy and total task completion time (in seconds) during the question-answering phase, and utilized the Post-Study System Usability Questionnaire (PSSUQ)~\cite{lewis1995ibm} to evaluate system usability and information quality, alongside the NASA Task Load Index (NASA-TLX)~\cite{hart1988development} to assess multi-dimensional cognitive load. Quantitative survey data was analyzed for statistical significance using the Wilcoxon signed-rank test~\cite{shier2004statistics}. We conducted semi-structured interviews centered around four open-ended questions: (1) \textit{Which condition better supported your understanding of the composite visualization, and why?} (2) \textit{Which approach was more cognitively demanding, and what specific factors caused this?} (3) \textit{Which specific features of \sysname were the most helpful?} and (4) \textit{What aspects of \sysname's interface or explanations were confusing or needed improvement?} Responses were transcribed and analyzed via thematic coding.

\subsubsection{Quantitative Results}

Given the non-parametric nature of our ordinal survey data and paired task times, all statistical comparisons between the two conditions were conducted using the Wilcoxon signed-rank test.

\textbf{Task Performance and Analytical Efficiency.} We first examined whether the medium affected the users' ability to successfully extract insights from the visualizations. We observed a ceiling effect for task accuracy across both conditions. Participants successfully answered the analytical questions regardless of whether they used the text baseline ($M_{\mathrm{base}} = 3.62, SD = 0.62$) or \sysname  ($M_{\mathrm{viz}} = 3.75, SD = 0.45$). This indicates that detailed text descriptions are sufficient for task completion if users invest the necessary time. However, the temporal cost differed drastically. Participants using \sysname ($M_{\mathrm{viz}}=194.38, SD=92.42$) were significantly faster than those relying on the text baseline ($M_{\mathrm{base}}=262.86, SD=150.79$), saving an average of 68 seconds, a statistically significant improvement ($p=0.039$) confirming that our \sysname accelerates insight extraction.

\begin{figure*}[t]
\centering  
\includegraphics[width=\textwidth]{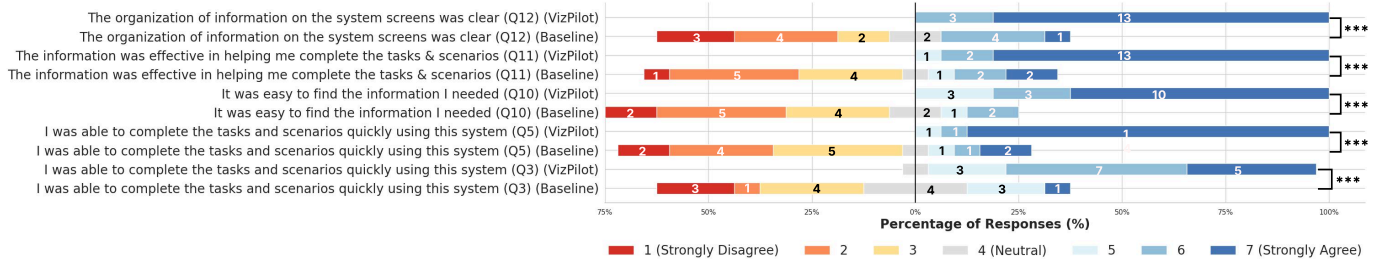}
\caption{\textbf{PSSUQ Results.} Results for selected PSSUQ items comparing the text baseline and \sysname. \sysname demonstrates a significant positive shift across both Information Quality Q10--Q12 and System Usefulness (Q3, Q5). Brackets with stars denote statistical significance based on the Wilcoxon signed-rank test (***: $p < 0.001$).}
\label{fig:pssuq}
\end{figure*}

 \textbf{System Usefulness and Information Quality.} {\color{black}We also examined participants' subjective perceptions of the onboarding experience using the Information Quality and System Usefulness dimensions of the PSSUQ. Participants reported significantly more favorable ratings for \sysname across all selected items. The largest differences were observed for locating referenced visual elements (Q10: $M_{\mathrm{viz}} = 6.44$, $\mathrm{SD} = 0.81$ vs.\ $M_{\mathrm{base}} = 3.06$, $\mathrm{SD} = 1.57$; $p = 0.0004$), organization of information (Q12: $M_{\mathrm{viz}} = 6.81$, $\mathrm{SD} = 0.40$ vs.\ $M_{\mathrm{base}} = 3.50$, $\mathrm{SD} = 2.10$; $p = 0.0005$), and perceived usefulness and learnability (Q3: $M_{\mathrm{viz}} = 6.00$, $\mathrm{SD} = 0.89$ vs.\ $M_{\mathrm{base}} = 3.44$, $\mathrm{SD} = 1.67$; $p = 0.0005$; Q5: $M_{\mathrm{viz}} = 6.81$, $\mathrm{SD} = 0.54$ vs.\ $M_{\mathrm{base}} = 3.38$, $\mathrm{SD} = 1.93$; $p = 0.0005$). These results suggest a strong user acceptance of \sysname and are consistent with the observed reduction in the task completion time. However, prior work has shown that subjective evaluations of AI-assisted systems can be influenced by expectation and novelty effects~\cite{10.1145/3529225}. We therefore interpret these findings primarily as evidence of perceived utility and acceptance, while relying on the objective performance measures (task accuracy and task completion time) reported above as stronger evidence of effectiveness.}

\begin{figure}[tb]
    \centering    \includegraphics[width=8cm]{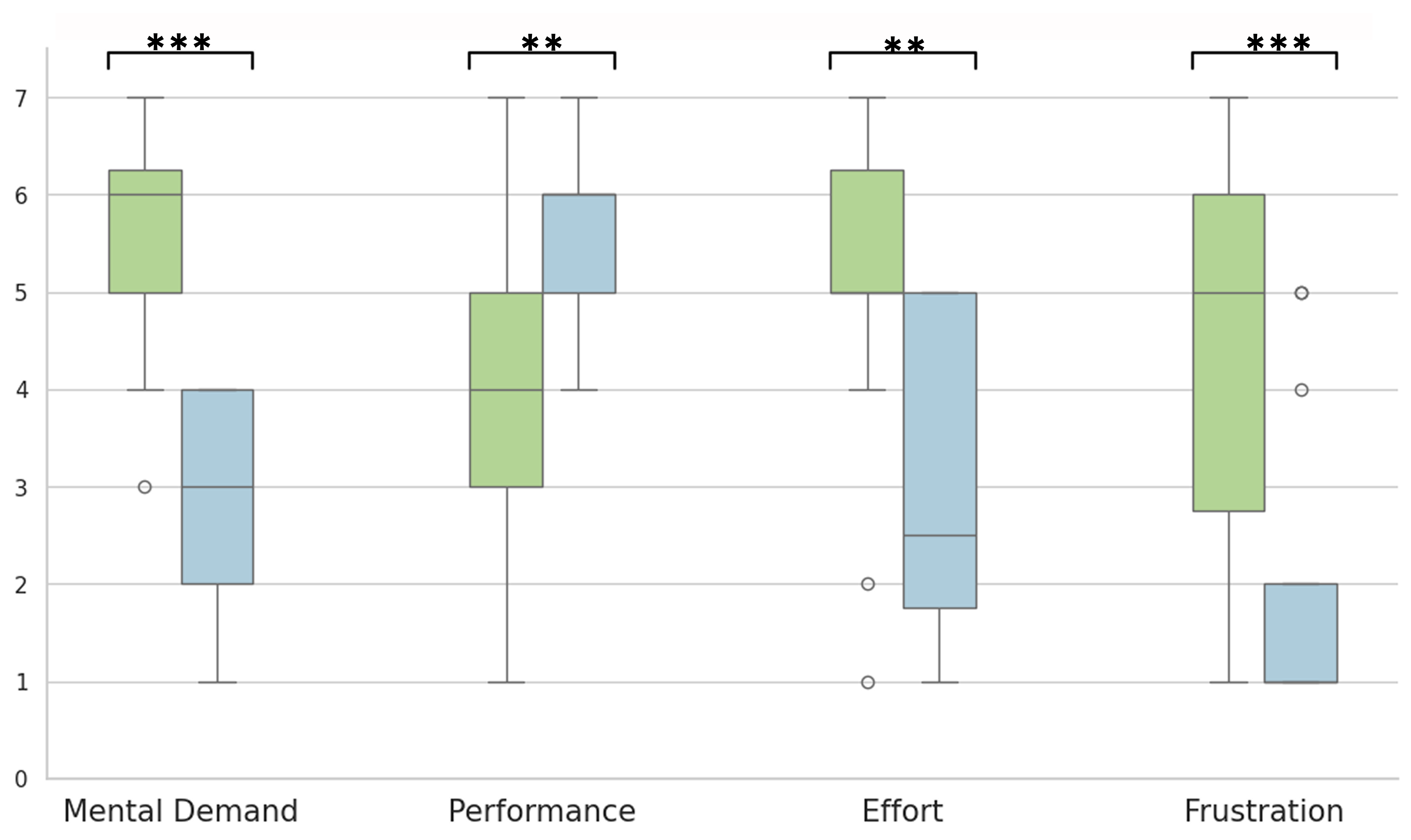} 
    \caption{\textbf{NASA-TLX Workload Assessment.} Box plots comparing the perceived workload across selected dimensions for the baseline (shown in green color) and \sysname (shown in blue color). Except for Performance, lower scores indicate a more favorable, lower-friction experience. \sysname significantly reduced mental demand, effort and frustration while improving perceived performance. Brackets with stars denote statistical significance based on the Wilcoxon signed-rank test (***: $p < 0.001$, **: $p < 0.01$).}
    \label{fig:nasa}
\end{figure}

\textbf{Cognitive Load.} {\color{black}To better understand the mechanisms underlying the observed efficiency gains, we analyzed participants' NASA-TLX workload ratings (Fig.~\ref{fig:nasa}). Participants reported significantly lower cognitive burden when using \sysname across several workload dimensions. The largest differences were observed in Mental Demand ($M_{\mathrm{viz}} = 2.94$, $\mathrm{SD} = 1.00$ vs.\ $M_{\mathrm{base}} = 5.62$, $\mathrm{SD} = 1.00$; $p = 0.001$) and Effort ($M_{\mathrm{viz}} = 2.94$, $\mathrm{SD} = 1.65$ vs.\ $M_{\mathrm{base}} = 5.19$, $\mathrm{SD} = 1.72$; $p = 0.007$). This result is consistent with the interpretation that the synchronized text and visual guidance reduced the need for participants to repeatedly switch between textual descriptions and visual elements, thereby mitigating split-attention effects during onboarding. Participants also reported substantially lower Frustration ($M_{\mathrm{viz}} = 1.88$, $\mathrm{SD} = 1.45$ vs.\ $M_{\mathrm{base}} = 4.31$, $\mathrm{SD} = 1.99$; $p = 0.001$) and higher perceived task Performance ($M_{\mathrm{viz}} = 5.62$, $\mathrm{SD} = 0.81$ vs.\ $M_{\mathrm{base}} = 3.94$, $\mathrm{SD} = 1.53$; $p = 0.003$) when using the onboarding by \sysname. These subjective workload assessments are consistent with the objective finding (task completion time) that participants completed the tasks more quickly with \sysname.}



\subsubsection{Qualitative Results}

To contextualize our quantitative findings, we conducted a thematic analysis of the post-study interviews. {\color{black}Rather than focusing on overall preference for the onboarding approaches, we analyzed participant feedback to identify specific interface mechanisms that influenced how users interpreted and navigated composite visualizations. The analysis revealed two advantages of the onboarding of \sysname in terms of cognitive support and onboarding structure, as well possible areas for future improvements..}

\textbf{Significant Mitigation of the Split-Attention Effect.} {\color{black} Participants consistently described the baseline condition as requiring frequent shifts between textual descriptions and visual elements. A recurring challenge was maintaining textual information in working memory while searching for the corresponding visual components.} As P7 noted, \textit{``In the process of reading the text, I need to constantly wander through the pictures and text... which makes me very tired.''} {\color{black}Participants attributed the reduced effort in the onboarding of \sysname primarily to the visual mapping provided by synchronized highlighting. These observations are consistent with the reductions in task completion time and perceived mental demand reported in the quantitative results.}

\textbf{Useful Dual-Mode Onboarding.} {\color{black}Almost all participants highlighted the dual-model onboarding provided by VizPilot. Participants noted that the step-by-step presentation helped them focus on individual concepts clearly. P12 described the experience as an \textit{``ordered step-by-step''} process, while P5 explained that fading irrelevant components \textit{``prevented me from getting overwhelmed''} by reducing visual clutter. Participants also appreciated the transition from guided onboarding to free exploration mode, which enabled them to revisit concepts and inspect details at their own pace. As P8 explained, they could \textit{``use the chat to check my understanding.''} These responses suggest that the combination of structured guidance and flexible exploration helped users navigate complex visualizations more systematically.}

\textbf{Areas for Improvement.} Participants also provided constructive feedback regarding our onboarding approach. A recurring theme was the desire for greater structural control. For example, P4 suggested making linear onboarding  \textit{``hierarchical''}, while P9 requested a \textit{``search feature for the guided tour''} to locate specific explanations. Users also suggested minor visual refinements, such as adding a  \textit{``blink''} before highlighting an element (P13) to catch subtle opacity changes. Participants also suggested for the chatbot to actively trigger highlights on components when answering.

%% file: source/Discussion.tex
\section{Discussion}

To contextualize our findings, we reflect on \sysname's deployment, design lessons, limitations, and future directions.

\subsection{Deployment and Generalizability}
\sysname is deployed as a browser extension that automatically extracts SVG structure and generates bitmap images to produce onboarding overlays, without the need of manual instrumentation. {\color{black}Our focus is on customized composite visualizations that still lack effective onboarding support, rather than commercial dashboards with built-in guidance or editorial visualizations designed for broad accessibility, which already provide comprehension affordances which are usually missing for customized composite visualizations. \sysname is a platform-agnostic approach and operates directly on rendered SVGs, making it compatible with all composite visualizations implemented by popular viusalization packages like D3.js and Vega-Lite, and broadly applicable across diverse web-based visualization ecosystems. \sysname assumes structurally grouped, DOM-accessible SVG elements. Canvas rendered visualizations and 
heavily flattened SVGs are outside the current scope and may result in a reduced component extraction accuracy. Interaction source code is optional but improves the interaction mapping when available.}

\subsection{Design Lessons}
A core challenge is reconciling the probabilistic nature of MLLMs with the deterministic demands of interactive interfaces. Directly binding explanations to visual elements risks hallucinated or imprecise mappings. \sysname addresses this via a hybrid design: MLLMs propose semantic mappings, which are then verified programmatically to ensure correctness. Beyond technical correctness, our findings suggest that effective onboarding also depends on how users build understanding over time. Drawing on narrative visualization principles~\cite{segel2010narrative}, a staged interaction design where by combining guided scrollytelling and free exploration, supports progressive mental model construction while maintaining user anatomy. Taken together, these findings position generative inference within constrained system-guided interaction workflows rather than as a standalone mechanism. 

\subsection{Limitations and Future Directions}
Despite the good performance of the onboarding by VizPilot, practical limitations remain. {\color{black}Our controlled study evaluated two visualizations selected for structural diversity, while the gallery evaluation provided broader coverage across all four design patterns. Future work should extend controlled studies to more visualization types and real-world sources. Furthermore, the subjective ratings of AI-enabled systems may be influenced by novelty and expectation effects \cite{10.1145/3529225}. We have mitigated this issue by including objective measures such as task accuracy and task completion time, but a more in-depth analysis of the potential bias introduced by expectation and novelty effects 
remains an avenue for future work.} \sysname currently assumes a single SVG container, limiting support for multi-view dashboards. Large SVGs may also exceed MLLM context limits, motivating targeted element selection. In addition, MLLM latency can hinder real-time interaction, making smaller or locally hosted models a promising direction. Future work also includes adaptive onboarding that adjusts pacing and detail based on users' expertise and interaction patterns.




%% file: source/conclusion.tex
\section{Conclusion}

As composite visualizations grow in complexity, the gap between developer intent and user comprehension widens. We present \sysname, an automated onboarding approach implemented as a browser extension comprising a Composite Visualization Analyzer, a two-stage MLLM-driven pipeline performing Semantic Inference and Semantic Mapping over SVG-based composite visualizations, and an Onboarding Interface supporting Narrative Scrollytelling and Free Exploration. Our evaluation demonstrates improvements in analytical efficiency, reduced cognitive load, and strong pipeline robustness across diverse visualization structures. This work illustrates how multi-stage MLLM outputs can be effectively integrated to deliver scalable, generalizable onboarding for web-based composite visualizations, lowering barriers for both developers and end-users.

\clearpage

%% file: source/appendix.tex
\appendix
\section{Appendices}
Section~\ref{subsec:semanticinference} details the Semantic Inference prompt, which reconstructs structured explanation units from multimodal inputs. Section~\ref{subsec:intermediate-repr} presents an example of the Intermediate Semantic Representation produced by the Semantic inference stage. Section~\ref{subsec:semantic-mapping} describes the Semantic Mapping prompt, which maps explanation units to SVG DOM elements. Section~\ref{subsec:mapped-repr} illustrates a complete Mapped Semantic Representation. {\color{black}Section~\ref{subsec:component-decomposition} provides the justification for the Visual Component Decomposition method. Section~\ref{subsec:quantitative-gallery} reports the metric definitions and the detailed results for each composite visualization for the quantitative evaluation of gallery visualizations. Section~\ref{subsec:expert-interview} describes about the expert interviews conducted with two experienced visualization researchers to assess \sysname’s utility in real-world authoring workflows,} and Section~\ref{subsec:modality-results} reports the results of the comparative analysis of input modality conditions.

\subsection{Semantic Inference}
\label{subsec:semanticinference}

The following prompt implements the semantic inference stage, where the model reconstructs a structured intermediate representation from multimodal inputs. It performs visual component decomposition, knowledge extraction, and semantic-to-text generation within a single constrained prompt to produce atomic explanation units.

\begin{tcolorbox}[
    colback=promptbg,
    colframe=promptframe,
    boxrule=0.5pt,
    arc=0pt,
    left=6pt, right=6pt, top=6pt, bottom=6pt,
    fontupper=\rmfamily\RaggedRight,
    breakable
]

You are an expert in data visualization analysis, visualization onboarding design, and composite visualization reasoning.
Your task is to analyze a composite visualization provided as: (1) an SVG specification, (2) a rendered bitmap image, (3) a user-authored description, and (4) the interaction source code, and produce a structured onboarding representation.

\medskip
\hrule
\medskip

\textbf{Structured Multi-Stage Reasoning (Strict)}\\[2pt]
You \textbf{must} follow a constrained three-stage reasoning pipeline. Each stage produces a structured intermediate representation that strictly bounds the next stage. You are \textbf{not} allowed to: skip stages, introduce new components after decomposition, or generate explanations not grounded in extracted knowledge.

\medskip
\hrule
\medskip

\textbf{Stage 1 - Visual Component Decomposition}\\[2pt]
Identify meaningful visual components using \textbf{both}: spatial layout cues from the image (position, alignment, adjacency) and structural grouping from the SVG. A base visual unit is the smallest visual structure that serves a distinct analytical role, which is distinguished by its own encoding, coordinate system, or data series, without decomposing to the level of individual marks.

\textbf{Rules:}
\begin{itemize}[nosep, topsep=2pt]
   \item A visual structure qualifies as a visual component \textbf{only if} it independently supports the extraction of at least one distinct form of visualization knowledge.
   \item Visual components can be hierarchical, where higher-level components contain multiple sub-components, provided each level independently yields extractable knowledge.
   \item Basic visual marks are \textbf{not} separate visual components, as they share redundant visual encodings and do not convey independent knowledge on their own.
   \item Visual components are categorized into:
     \begin{itemize}[nosep]
       \item \textbf{Data-driven components:} Coherent visual structures representing analytical data representations that require distinct, independent explanations.
       \item \textbf{Auxiliary components:} Contextual or reference elements (e.g., legends, axes, overview panels) that independently contribute extractable onboarding knowledge.
     \end{itemize}
\end{itemize}

\textbf{Rules:}



\textbf{Output}: \texttt{component\_id}, \texttt{role} (overview $|$ detail $|$ auxiliary $|$ shared), \texttt{chart\_type}, \texttt{approximate\_position}, \texttt{description}.

\medskip
\hrule
\medskip

\textbf{Stage 2 - Knowledge Extraction}\\[2pt]
For \textbf{each} visual component, extract structured knowledge using this taxonomy:

\begin{enumerate}[nosep, topsep=2pt, label=\arabic*)]
    \item \textbf{Structural} - Describe the overall composition and layout pattern (e.g., small multiples, coordinated views, overlays); explain shared elements (shared axes, legends) and how they connect different visual components.
    \item \textbf{Visual Encoding} - Explain how data attributes are mapped to visual channels (position, color, size, shape, or mark type); focus on how to read each individual mark within a visual component.
    \item \textbf{Data} - Identify what data is shown in each visual component: variables, units, categories, and time ranges.
    \item \textbf{Analytical} - Describe observable patterns, trends, comparisons, or relationships.
    \item \textbf{Interaction} - Infer supported interactions (hover, filter, brush, zoom) and explain what analytical tasks they enable.
\end{enumerate}

\textbf{Constraints:}
\begin{itemize}[nosep, topsep=2pt]
    \item Knowledge \textbf{must} be grounded in visible evidence
    \item Analytical knowledge \textbf{must} describe observable patterns
    \item Interaction knowledge \textbf{must} only be included if supported by the interaction source code or visual evidence
    \item No new components may be introduced
\end{itemize}

\medskip
\hrule
\medskip

\textbf{Stage 3 - Semantic-to-Text Generation}\\[2pt]
Transform structured knowledge into atomic explanation units. Each unit \textbf{must}:
\begin{itemize}[nosep, topsep=2pt]
    \item correspond to \textbf{one} visual component
    \item correspond to \textbf{one} knowledge category
    \item express \textbf{one} atomic concept
    \item derive directly from Stage~2 knowledge
\end{itemize}

\medskip
\hrule
\medskip

\textbf{Use of Interaction Source Code}\\[2pt]
The Interaction source code may be used \textbf{only} to:
\begin{itemize}[nosep, topsep=2pt]
    \item Infer the existence and intent of user interactions (e.g., hover, click, brush, zoom, filter, linked highlighting)
    \item Understand interaction scope across multiple visual components (e.g., coordinated or linked components)
    \item Disambiguate interactions that are not visually obvious from the static SVG or rendered image alone
\end{itemize}

\medskip
\hrule
\medskip

\textbf{Traceability Requirement (Critical)}\\[2pt]
Each \texttt{explanation\_step} \textbf{must} be traceable to a visual component (\texttt{target\_component\_id}), a knowledge category (\texttt{step\_type}), and a specific extracted knowledge unit. If it cannot be traced, \textbf{do not} generate it.

\medskip
\hrule
\medskip

\textbf{Strategic Narrative Rules}
\begin{itemize}[nosep, topsep=2pt]
    \item \textit{Relational Logic:} Explain how visual components relate (e.g., filtering, shared axes)
    \item \textit{Dual-Encoding Clarity:} Explain combined encodings clearly
    \item \textit{Action-Oriented Language:} Guide user attention (e.g., ``Look at the colored dots'')
    \item Include at least one analytical ``Aha'' insight
    \item Use mark-level terms: ``colored dot'', ``bar'', ``line'', ``box''
\end{itemize}

\textbf{Conceptual Framing Rule (Critical):} Start by establishing (1) overall goal of the visualization, (2) main visual components, (3) role of each component. Do \textbf{not} begin with mark-level explanations.

\medskip

\textbf{Cross-Component Coordination Rule:} Include at least \textbf{two} steps explaining relationships between visual components (e.g., overview $\rightarrow$ detail navigation, filtering, highlighting, shared axes, causal relationships).

\medskip

\textbf{Atomicity \& Cognitive Load:} Each step must be 1--2 sentences ($\approx$15--35 words) expressing only \textbf{one} idea. Recommended progression: structure $\rightarrow$ data $\rightarrow$ encodings $\rightarrow$ analytical $\rightarrow$ interactions.

\medskip
\hrule
\medskip

\textbf{Output Format (JSON only):}
\begin{lstlisting}[basicstyle=\ttfamily\small, breaklines=true, columns=flexible]
{
  "explanation_steps": [
    {
      "step_type": "structure | encoding | data | analytical | interaction",
      "target_component_id": "...",
      "priority": 1,
      "explanation_text": "..."
    }
  ]
}
\end{lstlisting}

\medskip
\hrule
\medskip

\textbf{Quality Requirements}
\begin{itemize}[nosep, topsep=2pt]
    \item Include at least one step for: structure, data, encoding, analytical
    \item Include interaction steps only if supported by interaction source or visual evidence
    \item Include at least one workflow guidance step
    \item Include at least \textbf{two} cross-component coordination steps
    \item \texttt{explanation\_text} must be concise and user-facing
    \item No hallucinated elements or interactions
\end{itemize}

\textbf{Important:}
\begin{itemize}[nosep, topsep=2pt]
    \item Each explanation step must be traceable to a visual component and knowledge type
    \item Do \textbf{not} output intermediate reasoning
    \item Return \textbf{JSON only}
\end{itemize}

\medskip
\hrule
\medskip

\textbf{Input}\\[2pt]
The following will be provided: (1) user-authored description, (2) SVG code, (3) rendered bitmap image, (4) interaction source code.

\end{tcolorbox}

\subsection{Intermediate Semantic Representation}
\label{subsec:intermediate-repr}

To illustrate the output of the Semantic Inference stage, we present a
single representative explanation unit from the structured explanation steps array generated for the MatrixWave
visualization (Fig.~\ref{fig:onboardingmodes}) (Listing~\ref{lst:semantic-inference-output}). Each entry encodes a single
atomic concept tied to one visual component and one knowledge category, ready
to be consumed by the Semantic Mapping stage. The \texttt{step\_type}
field classifies the knowledge category. The
\texttt{target\_component\_id} field identifies the logical component within
the composite visualization to which the explanation unit belongs. The \texttt{priority} field governs
presentation order.

\begin{lstlisting}[language=json, caption={Example JSON output for a single explanation unit produced by Semantic Inference stage.}, label={lst:semantic-inference-output}]
{
  "step_type": "structure",
  "target_component_id": "step_bars_shared",
  "priority": 2,
  "explanation_text":"Use the slim header bars as the overview. They summarize per-page traffic at each step and align with the rows"
}
\end{lstlisting}

\subsection{Semantic Mapping}
\label{subsec:semantic-mapping}

The following prompt implements the semantic mapping stage, which maps abstract explanation units to SVG elements. It employs hierarchical selector inference to construct deterministic selector paths from component-level containers to encoding-level elements, enabling precise and verifiable mappings.

\begin{tcolorbox}[
    colback=promptbg,
    colframe=promptframe,
    boxrule=0.5pt,
    arc=0pt,
    left=6pt, right=6pt, top=6pt, bottom=6pt,
    fontupper=\rmfamily\RaggedRight,
    breakable
]

You are an expert in SVG structure analysis, DOM reasoning, and visualization semantic mapping.
Your task is to map each explanation unit to exact SVG elements using precise, hierarchical, and verifiable SVG selectors.

\medskip
\hrule
\medskip

\textbf{Hierarchical Selector Inference (Core Method)}\\[2pt]
You \textbf{must} construct selector paths that reflect the DOM hierarchy:

\medskip
\texttt{parent-container > intermediate-group > target-element}
\medskip

Rules:
\begin{itemize}[nosep, topsep=2pt]
    \item Always begin from the resolved component container
    \item Progressively refine through nested \texttt{<g>} groups toward target marks
    \item Flat selectors (e.g., \texttt{rect} alone) are \textbf{not} acceptable unless uniquely identifying
    \item Do \textbf{not} invent IDs, classes, or attributes — use only those present in the SVG
\end{itemize}

\medskip
\hrule
\medskip

\textbf{Two-Stage Mapping Pipeline}

\medskip

\textbf{Stage 1 - Component-Level Mapping (Container Resolution)}\\
Identify the SVG container (\texttt{<g>}, \texttt{<svg>}, or grouping node) that bounds each visual component.
Use spatial layout cues (left/right/top/bottom), \texttt{<g>} nesting depth, and structural grouping to resolve the correct container.

\medskip

\textbf{Stage 2 - Encoding-Level Mapping (Mark Resolution)}\\
Within the resolved container \textbf{only}, identify the SVG elements that correspond to the semantic unit of each explanation step.
These include marks such as bars (\texttt{rect}), lines (\texttt{path}/\texttt{line}), dots (\texttt{circle}), areas, or glyphs.
Encoding selectors \textbf{must} be scoped within their corresponding component container.

\medskip
\hrule
\medskip

\textbf{Selector Construction Rules}
\begin{itemize}[nosep, topsep=2pt]
    \item Prefer \textbf{IDs} $>$ \textbf{classes} $>$ attribute selectors $>$ structural selectors
    \item If multiple attributes exist, include \textbf{both} for precision
    \item Return \textbf{multiple candidate selectors} per category when ambiguity exists
\end{itemize}

\medskip
\hrule
\medskip

\textbf{Ambiguity Resolution}\\[2pt]
If multiple candidates exist, return all with confidence scores:
\begin{itemize}[nosep, topsep=2pt]
    \item \textbf{high} - strong structural and visual evidence
    \item \textbf{medium} - partial evidence
    \item \textbf{low} - weak or ambiguous match
\end{itemize}
Use component hierarchy, relative positioning, and repetition patterns to disambiguate.

\medskip
\hrule
\medskip

\textbf{Output Format (JSON only):}
\begin{lstlisting}[basicstyle=\ttfamily\small, breaklines=true, columns=flexible]
{
  "mappings": [
    {
      "step_priority": 1,
      "target_component_id": "...",
      "step_type": "structure | encoding | data | analytical | interaction",
      "component": {
        "candidates": [
          {
            "selector": "...",
            "confidence": "high | medium | low",
            "reason": "..."
          }
        ]
      },
      "encoding": {
        "candidates": [
          {
            "selector": "...",
            "confidence": "high | medium | low",
            "reason": "..."
          }
        ]
      }
    }
  ]
}
\end{lstlisting}

\medskip
\hrule
\medskip

\textbf{Important:}
\begin{itemize}[nosep, topsep=2pt]
    \item Each mapping entry corresponds to \textbf{exactly one} explanation unit, identified by \texttt{step\_priority}, \texttt{target\_component\_id}, and \texttt{step\_type}
    \item Encoding selectors \textbf{must} be scoped within their resolved component container
    \item Do \textbf{not} output executable code or explanations outside the JSON
    \item Selectors must be deterministic and verifiable against the SVG DOM
    \item Return \textbf{JSON only}
\end{itemize}

\medskip
\hrule
\medskip

\textbf{Input}\\[2pt]
The following will be provided: (1) Explanation units from semantic inference, (2) SVG code

\end{tcolorbox}


\subsection{Mapped Semantic Representation}
\label{subsec:mapped-repr}

To demonstrate \sysname's output after the Semantic Mapping stage, we provide a representative Mapped Explanation 
Unit generated for the MatrixWave visualization (Fig.~\ref{fig:onboardingmodes}) (Listing~\ref{lst:json-output}). The output follows a structured format 
where the \texttt{step\_type} dictates the knowledge category. The 
\texttt{target\_component} field identifies the logical parent visual component within the composite 
visualization, providing the spatial context necessary during the 
Semantic Mapping stage. The \texttt{resolvedSelector} represents the 
final, validated DOM path produced after the MLLM output undergoes the programmatic validation step. The 
\texttt{actions} array tells the 
\sysname frontend how to visually emphasize the corresponding SVG 
elements. This structured output ensures that the onboarding remains 
interactive and synchronized with the live visualization state.

\begin{lstlisting}[language=json, caption={Example JSON output for a single mapped explanation unit produced by the Semantic Mapping stage.}, label={lst:json-output}]
{
  "step_id": "step_2",
  "step_type": "structure",
  "text": "Use the slim header bars as the overview..",
  "target_component": "step_bars_shared",
  "resolvedSelector": "g.header g.mathead rect.val",
  "actions": ["highlight"]
}
\end{lstlisting}


{\color{black}
\subsection{Visual Component Decomposition Justification}
\label{subsec:component-decomposition}
\sysname performs decomposition at the level where the visual structure should independently support the extraction of at least one distinct form of visualization knowledge rather than at the level of individual visual marks. We justify this design choice as follows.

\textbf{Alignment with the onboarding task.} \sysname is designed to help first time users build a structural understanding of an unfamiliar composite visualization as what components exist, how they relate, and what each contributes. This requires identifying visual components not enumerating every mark within them. A user does not need an onboarding step explaining each individual bar in a bar chart as they need to understand that the bar chart exists, what it encodes, and 
how it relates to adjacent components. Mark level detail is conveyed within a component's explanation, not by further decomposing the chart into per mark components.

\textbf{Avoiding combinatorial explosion.} Composite visualizations in our gallery contain charts with tens to hundreds of individual marks. Treating each mark as a 
separate component would produce an unusable number of onboarding steps, which would be against the purpose of structured onboarding and overwhelming the user with redundant, near identical explanations.

\subsubsection*{Illustrative Example}

\begin{figure}[tb]
    \centering
\includegraphics[width=\columnwidth]{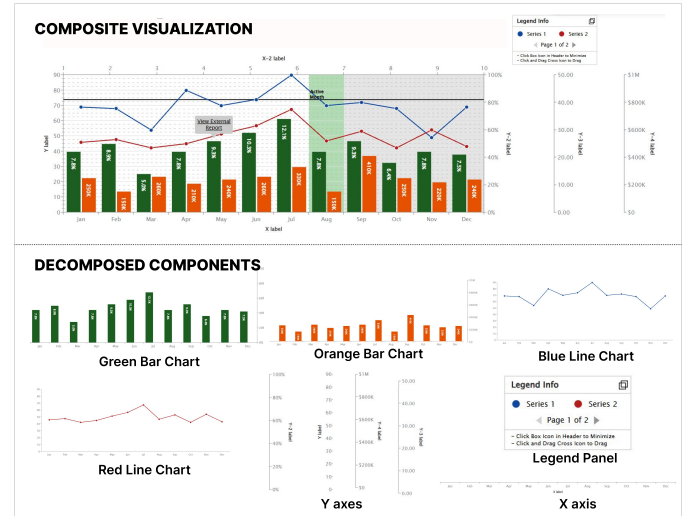} 
    \caption{Component decomposition of a composite visualization from ZingChart~\cite{zingchart_labelmania}.}
    \label{fig:compdecomp}
\end{figure}

Figure~\ref{fig:compdecomp} illustrates how our knowledge driven decomposition criterion operates in practice on a composite visualization. This example demonstrates the important boundary decisions that follow directly from our definition of a visual component:

\textbf{Data-driven components within a shared coordinate space.} Although the ``green bar chart'' and ``orange bar chart'' are rendered within the same main plot area and share an X-axis, they are treated as two distinct data-driven visual components. They encode different data dimensions using distinct mark types and colors, each contributing independent visualization knowledge that requires separate explanation. Similarly, the two overlaying line charts are treated as separate data-driven components. Merging all four into a single ``main plot'' visual component would be against the fact that they encode distinct variables and require separate onboarding explanations.

\textbf{Exclusion of individual visual marks.} Despite the chart containing individual green bars, orange bars, and line points, none of these individual marks constitute separate visual components because they share redundant visual encoding knowledge. A first time user needs to understand what the green bar series represents as a whole across months, rather than receiving near identical explanations per bar.

\textbf{Auxiliary visual components.} X-axis, Y-axes, and legend panel are each treated as separate visual components because they serve meaningfully distinct onboarding purposes, where axes provide scale context and the legend provides series identification. Each warrants its own explanation step rather than being subsumed into a generic ``main chart'' component.

\subsection{Quantitative Evaluation of Gallery Visualizations: Metrics and Detailed Results for Each Composite Visualization}
\label{subsec:quantitative-gallery}

This section provides full definitions of the evaluation metrics, along with the detailed results of each composite visualization.

\subsubsection*{Metric Definitions}

\textbf{Decomposition (Precision, Recall, F1).} Precision measures the proportion of system-identified visual components that correspond to a ground-truth visual component, Recall measures the proportion of ground-truth visual components our pipeline correctly identified and F1 is their harmonic mean. Ground-truth visual component lists were established by the authors prior to pipeline execution using the Visual Component Decomposition definition.

\textbf{Semantic Mapping (Exact, Partial).} For each generated onboarding step, this dimension evaluates whether the generated SVG selector correctly maps the visual region described in the onboarding step. Exact accuracy counts only fully correct selectors and Partial accuracy additionally counts near misses such as a selector targeting a parent container one level too broad.

\textbf{Explanation Coherence (Type alignment,
Component alignment, Atomicity).} Each step is assessed on three independent sub-criteria: Type alignment, which evaluates whether the explanation matches its declared step type (structure, data, encoding, analytical, or interaction); Component alignment, which evaluates whether the text describes content belonging to its target visual component; and Atomicity, which checks whether the step expresses exactly one knowledge category without mixing structural, encoding, data, analytical, or interaction knowledge. Each sub-criterion is reported as a pass rate across all steps for a visualization.

\textbf{End-to-End Success Rate (E2E).} A ground-truth visual component counts as a full success only if it was correctly decomposed, every step generated for it received a correct or partial semantic mapping judgment, and every step passed all three explanation coherence criteria. The E2E rate is the proportion of ground-truth visual components satisfying this joint criterion, making missed components explicitly visible in the score.

\subsubsection*{Detailed Results for Each Composite Visualization}

Tables~\ref{tab:full_results_dim} and~\ref{tab:full_results_e2e} report detailed results for each composite visualization across all 18 gallery visualizations for each evaluation dimension and the end-to-end success rate, respectively.




\begin{table*}[t]
\centering
\caption{Detailed results for each visualization across decomposition, semantic mapping, and coherence dimensions.}
\label{tab:full_results_dim}
\begin{tabular}{l l ccc cc ccc}
\toprule
\multirow{2}{*}{\textbf{Composite Pattern}}
  & \multirow{2}{*}{\textbf{Visualization}}
  & \multicolumn{3}{c}{\textbf{Decomposition}}
  & \multicolumn{2}{c}{\textbf{Semantic Mapping}}
  & \multicolumn{3}{c}{\textbf{Coherence}} \\
\cmidrule(lr){3-5} \cmidrule(lr){6-7} \cmidrule(lr){8-10}
  & & Prec. & Rec. & F1 & Exact & Partial & Type & Comp & Atom \\
\midrule
\multirow{6}{*}{Juxtaposition} 
  & PrettiSmart~\cite{wen2025prettismart}             & 1.000 & 1.000 & 1.000 & 0.940 & 1.000 & 0.889 & 1.000 & 1.000 \\
  & Diffseer~\cite{10.1109/MCG.2023.3248289}          & 1.000 & 1.000 & 1.000 & 0.843 & 0.929 & 1.000 & 1.000 & 0.930 \\
  & RuleMatrix~\cite{rulematrix}                      & 0.889 & 1.000 & 0.941 & 0.929 & 0.929 & 1.000 & 0.929 & 1.000 \\
  & QuantumEyes~\cite{10319321}                       & 0.909 & 0.909 & 0.909 & 0.950 & 1.000 & 0.950 & 1.000 & 0.800 \\
  & Multiple Time Series~\cite{zingchart_multiscale}  & 1.000 & 1.000 & 1.000 & 0.917 & 1.000 & 1.000 & 1.000 & 0.934 \\
  & Label Mania~\cite{zingchart_labelmania}           & 0.917 & 0.846 & 0.880 & 1.000 & 1.000 & 0.944 & 1.000 & 1.000 \\
\midrule
\multirow{5}{*}{Overloading}
  & MatrixWave~\cite{zhao2015matrixwave}              & 1.000 & 0.850 & 0.919 & 0.882 & 1.000 & 0.941 & 0.941 & 0.824 \\
  & VIStory~\cite{vistory}                            & 1.000 & 0.800 & 0.889 & 0.900 & 1.000 & 1.000 & 1.000 & 0.900 \\
  & AttentionFlow~\cite{10.1145/3437963.3441703}      & 1.000 & 1.000 & 1.000 & 0.917 & 1.000 & 1.000 & 0.917 & 0.933 \\
  & EgoLines~\cite{zhao2016egocentric}                & 0.818 & 0.900 & 0.857 & 0.923 & 0.923 & 1.000 & 1.000 & 0.792 \\
  & Dodrio~\cite{wang-etal-2021-dodrio}               & 1.000 & 0.850 & 0.919 & 0.850 & 0.850 & 1.000 & 1.000 & 0.875 \\
\midrule
\multirow{3}{*}{Nesting}
  & PEARL~\cite{pearl}                                & 0.875 & 0.875 & 0.875 & 0.938 & 0.938 & 1.000 & 0.875 & 1.000 \\
  & PonziLens+~\cite{10794804}                        & 1.000 & 1.000 & 1.000 & 0.936 & 1.000 & 0.882 & 1.000 & 0.765 \\
  & UpSet~\cite{6876017}                              & 1.000 & 0.833 & 0.909 & 0.900 & 1.000 & 1.000 & 1.000 & 0.900 \\
\midrule
\multirow{4}{*}{Superimposition}
  & CandleStick on Mobile~\cite{echarts_candlestick_on_mobile} & 1.000 & 0.889 & 0.941 & 0.886 & 0.929 & 0.929 & 1.000 & 0.929 \\
  & Shanghai Index~\cite{echarts_shanghai_index}       & 1.000 & 1.000 & 1.000 & 0.857 & 0.929 & 0.929 & 1.000 & 1.000 \\
  & A Timeline of Media-Inflamed Fears~\cite{zingchart_timeline} & 1.000 & 1.000 & 1.000 & 0.929 & 1.000 & 0.929 & 1.000 & 0.929 \\
  & Mixed Bar and Staggered Line~\cite{zingchart_mixedbar} & 1.000 & 0.878 & 0.935 & 1.000 & 1.000 & 1.000 & 1.000 & 0.917 \\
\midrule
\textit{Mean} & & 0.967 & 0.918 & 0.940 & 0.912 & 0.968 & 0.966 & 0.981 & 0.913 \\
\bottomrule
\end{tabular}
\end{table*}

\begin{table}[t]
\centering
\caption{End-to-End Success Rate for each visualization across all 18 gallery visualizations.}
\label{tab:full_results_e2e}
\begin{tabular}{p{0.5\columnwidth} c}
\toprule

\textbf{Visualization} & \textbf{End-to-End Success Rate} \\

\midrule

PrettiSmart~\cite{wen2025prettismart}      & 1.000 \\
Diffseer~\cite{10.1109/MCG.2023.3248289}       & 1.000 \\
RuleMatrix~\cite{rulematrix}     & 0.800 \\
QuantumEyes~\cite{10319321}    & 0.909 \\
Multiple Time SeriesMultiple Time Series~\cite{zingchart_multiscale}      & 1.000 \\
Label Mania~\cite{zingchart_labelmania}     & 0.846 \\
MatrixWave~\cite{zhao2015matrixwave}     & 0.850 \\
VIStory~\cite{vistory}        & 0.800 \\
AttentionFlow~\cite{10.1145/3437963.3441703}  & 0.875 \\
EgoLines~\cite{zhao2016egocentric}       & 0.800 \\
Dodrio~\cite{wang-etal-2021-dodrio}          & 0.800 \\
PEARL~\cite{pearl}          & 0.889 \\
PonziLens+~\cite{10794804}        & 1.000 \\
UpSet~\cite{6876017}          & 0.833 \\
CandleStick on Mobile~\cite{echarts_candlestick_on_mobile}     & 0.889 \\
Shanghai Index~\cite{echarts_shanghai_index}       & 1.000 \\
A Timeline of Media-Inflamed Fears~\cite{zingchart_timeline}         & 1.000 \\
Mixed Bar and Staggered Line~\cite{zingchart_mixedbar}       & 0.878 \\

\midrule

\textit{Mean} & 0.898 \\

\bottomrule
\end{tabular}
\end{table}

Across all 18 visualizations, \sysname achieves strong and consistent performance on all three dimensions. Decomposition precision consistently exceeds recall, indicating our approach more often misses a minor visual component than invents a phantom one, with lower recall cases concentrated in visualizations with densely overlapping components or ambiguous spatial layouts. Semantic mapping partial accuracy is uniformly high, with lower exact accuracy reflecting cases where \sysname selects a slightly broader parent container rather than the precise target element. Among coherence sub-criteria, component alignment is the strongest across all visualizations while atomicity is the weakest, reflecting occasional steps that blend encoding and analytical knowledge within a single explanation unit. The end-to-end success rate, which jointly requires correct decomposition, grounding, and explanation quality, further confirms that errors are not dominated by a single failure mode but are instead distributed sparsely across cases. Overall, results suggest that \sysname generalizes effectively across heterogeneous composite design patterns.

\subsection{Expert Interview}
\label{subsec:expert-interview}
To assess \sysname's utility in real-world authoring workflows, we conducted an evaluation with two visualization researchers with more than 7 years of experience in visualization development. Both experts noted that manually building onboarding materials typically requires labor-intensive documentation or custom video rendering. After testing \sysname, they confirmed the automated pipeline accurately captured the structural intent, visual encodings, and high-level patterns. They highlighted that \sysname drastically reduces developer effort by eliminating the need to manually code event listeners, while delivering a significantly more engaging end-user experience than static text. Both expressed a strong willingness to integrate the tool into their regular workflows, suggesting future enhancements such as animated state transitions for complex interaction chains and improved accuracy for the QA assistant on high-level analytical queries.} 

\subsection{Results of Comparative Analysis of Input Modalities}
\label{subsec:modality-results}

{\color{black}
We present inference outputs across four input modality conditions for four composite visualizations: PrettiSmart~\cite{wen2025prettismart}, RuleMatrix~\cite{rulematrix}, PonziLens~\cite{10794804}, and Shanghai Index~\cite{echarts_shanghai_index}, spanning distinct composite design patterns. Because interaction source code is explicitly restricted to interaction inference in our pipeline design, structural and encoding outputs are identical between SVG\,+\,Image and the full pipeline. Therefore, Tables~\ref{tab:merged-decomp} and~\ref{tab:merged-encoding} compare three conditions (SVG Only, Image Only, SVG\,+\,Image), while Table~\ref{tab:merged-interaction} compares all four to isolate the specific contribution of interaction source code.

\begin{table*}[ht]
\centering
\caption{Visual Component decomposition across four visualizations and three modality conditions.}
\label{tab:merged-decomp}
\renewcommand{\arraystretch}{1.4}
\small
\begin{tabularx}{\linewidth}{p{0.11\linewidth} X X X}
\toprule
\textbf{Visualization}
  & \textbf{SVG Only}
  & \textbf{Image Only}
  & \textbf{SVG\,+\,Image} \\
\midrule

PrettiSmart &
11 geometry-driven layers; layer-driven naming
(e.g., \texttt{flow\_links\_layer},
\texttt{simulations\_marks\_layer}); structural step
describes geometric relationships without spatial grounding:
``Start at the top with Simulation strips \ldots\ read central views
where rectangles, curved links, and colored areas align to the
Contract line \ldots'' &
5 coarser components; appearance-driven naming
(e.g., \texttt{flow\_diagram},
\texttt{simulation\_thumbnails}); structural step uses spatial
language: ``The large central blue-pink flow between Contract and
addresses is the primary view. Small heatmaps above and the state
strip below provide overview and temporal context.'' &
Spatially grounded naming with selector-ready identifiers
(e.g., \texttt{sim0\_heatmap}--\texttt{sim7\_heatmap},
\texttt{flow\_paths\_layer}); structural step combines spatial
and structural language: ``Start at the top row of eight
small-multiple cards \ldots\ move down to the large middle view
where colored curved paths link the Contract row to addresses
at the same steps \ldots'' \\

\midrule

RuleMatrix &
15 steps; precise layer-driven component names
(e.g., \texttt{g.rules}, \texttt{g.mo-fidelity}); selectors
resolve to specific DOM elements including transform-indexed
histogram columns. &
Component-level mapping only (no step list); selectors
reasonable (e.g., \texttt{g.rm-labels}, \texttt{g.func})
but unanchored to step text; interaction inference absent. &
14 steps; spatially grounded names
(e.g., \texttt{output\_probability}, \texttt{fidelity\_gauge});
selectors confirmed against both structure and appearance. \\

\midrule

PonziLens &
12 steps; all grounded to \texttt{g.overview\_G}; specific
encoding candidates (loop bracket paths, action dot colors);
all three modules (Path Feature, Path Grouping, Execution
Detail) correctly identified. &
14 steps but all selectors low-confidence generic
(\texttt{svg}, \texttt{g}, \texttt{circle}); specialized
domain vocabulary (Ponzi features, blockchain execution)
cannot be anchored to SVG primitives without structural
context; zero DOM actionability. &
17 steps; most specific outputs across all four
visualizations
(e.g., \texttt{g.legend g.WRITE circle},
\texttt{g.overview\_G path[stroke="\#9b8ea9"~i]});
domain-specific component roles recovered via joint
SVG\,+\,Image reasoning. \\

\midrule

Shanghai Index &
11 steps; candlestick color-attribute selectors are strong;
some selectors anchored to exact \texttt{d} attribute values;
MA lines correctly identified by stroke color. &
18 steps but almost entirely low-confidence generic selectors
(\texttt{g[clip-path]}, \texttt{g[cursor="pointer"]});
strong visual semantic reasoning but zero DOM actionability;
most extreme semantic-grounding gap across all tested
visualizations. &
14 steps; cleanest and most specific selector outputs
(e.g., \texttt{g[clip-path="url(\#zr0-c0)"]} \texttt{> path[stroke="rgb(84,112,198)"~i]});
image confirms component boundaries that SVG-only
over-specifies; fragile SVG-only selectors resolved. \\

\bottomrule
\end{tabularx}
\end{table*}

\textbf{Key Findings (Table~\ref{tab:merged-decomp}).}
SVG-only consistently produces DOM-valid selectors with structural precision but shallow spatial semantics and its naming is geometry driven and its structural narratives describe geometric relationships rather than spatial reading strategies. Image-only produces richer spatial language and coarser but more semantically meaningful component groupings, but critically relies on \texttt{target\_description} fields rather than \texttt{component\_id}s, making all outputs unresolvable by the semantic mapping stage. SVG\,+\,Image combines both, where spatially grounded naming with selector ready identifiers. This pattern holds consistently across all four visualizations, with PonziLens showing the most severe image-only degradation as its specialized domain vocabulary (Ponzi features, execution modules) cannot be anchored to generic SVG primitives without structural context, suggesting that domain complexity amplifies the image-only limitation.

\begin{table*}[ht]
\centering
\caption{Selected encoding steps across four visualizations and three modality conditions.}
\label{tab:merged-encoding}
\renewcommand{\arraystretch}{1.4}
\small
\begin{tabularx}{\linewidth}{p{0.11\linewidth} p{0.13\linewidth} X X X}
\toprule
\textbf{Visualization}
  & \textbf{Mark}
  & \textbf{SVG Only}
  & \textbf{Image Only}
  & \textbf{SVG\,+\,Image} \\
\midrule

PrettiSmart
& Function call rectangles &
Brown\,=\,Payable; beige\,=\,NonPayable.
\newline\textit{Selector: component-level} &
``Tan squares mark function call events.''
\newline\textit{No DOM anchor} &
Brown square\,=\,Payable call; beige\,=\,NonPayable call,
placed at the exact step on the corresponding row.
\newline\textit{Selector: component-grounded, appearance confirmed} \\

& Flow curves &
Pink\,=\,Money Out; blue\,=\,Money In; thicker\,=\,larger amount.
\newline\textit{Selector: component-level} &
``Curved strokes with arrowheads indicate money in or out.''
\newline\textit{No DOM anchor} &
Pink path\,=\,negative outflow; blue path\,=\,positive inflow;
origin and landing point identify sender and receiver.
\newline\textit{Selector: direction semantics visually confirmed} \\

\midrule

RuleMatrix
& Rule–output cells &
Cell color encodes output class probability; cell size encodes
support count.
\newline\textit{Selector: \texttt{g.rules} container} &
``Colored grid cells show rule strength.''
\newline\textit{No DOM anchor; no size encoding mentioned} &
Color encodes class probability; size encodes support; both
confirmed against rendered cell appearance.
\newline\textit{Selector: \texttt{g.rules rect} — appearance confirmed} \\

& Fidelity gauge &
Arc length encodes fidelity score.
\newline\textit{Selector: \texttt{g.mo-fidelity}} &
``A gauge-like arc shows model accuracy.''
\newline\textit{No DOM anchor} &
Arc length and fill color both encode fidelity; full arc
confirmed visually.
\newline\textit{Selector: \texttt{g.mo-fidelity path} — confirmed} \\

\midrule

PonziLens
& Loop bracket paths &
Bracket shape encodes loop structure; color encodes action type.
\newline\textit{Selector: \texttt{g.overview\_G path}} &
``Arcs connect related events.''
\newline\textit{Generic anchor; action-type color not mentioned} &
Bracket shape\,=\,loop; color\,=\,action type (READ, WRITE,
CALL); confirmed against rendered paths.
\newline\textit{Selector: \texttt{g.overview\_G path[stroke="\#9b8ea9"~i]}} \\

& Action dots &
Dot color encodes action type (READ, WRITE, CALL).
\newline\textit{Selector: \texttt{g.legend g.WRITE circle}} &
``Colored dots indicate event types.''
\newline\textit{No anchor; WRITE/READ/CALL distinction absent} &
Dot color encodes action type; legend confirms READ/WRITE/CALL
mapping visually.
\newline\textit{Selector: \texttt{g.legend g.WRITE circle} — confirmed} \\

\midrule

Shanghai Index
& Candlestick marks &
Color (red/green) encodes price direction; body height encodes
open-close range.
\newline\textit{Selector: color-attribute selector — strong} &
``Red and green bars show price movement.''
\newline\textit{No DOM anchor; wick encoding absent} &
Red\,=\,price decline; green\,=\,price rise; wick height
encodes high-low range; both confirmed.
\newline\textit{Selector: \texttt{path[stroke="rgb(84,112,198)"~i]} — confirmed} \\

& MA lines &
Stroke color distinguishes MA5, MA10, MA20.
\newline\textit{Selector: anchored by stroke color — correct} &
``Colored lines show moving averages.''
\newline\textit{MA5/MA10/MA20 distinction absent; no anchor} &
Stroke color confirmed per MA series; clip-path container
correctly disambiguates series.
\newline\textit{Selector: \texttt{g[clip-path]} \texttt{> path} — confirmed} \\

\bottomrule
\end{tabularx}
\end{table*}

\textbf{Key Findings (Table~\ref{tab:merged-encoding}).}
Color semantics and mark-type meanings are recovered consistently across SVG-only and SVG\,+\,Image for all four visualizations. The critical difference is mapping, where image-only descriptions rely on visual anchors (e.g., ``tan squares'', ``colored dots'', ``arc shows accuracy'') that cannot be resolved to any DOM node, making them unactionable for selector inference regardless of how semantically accurate the description is. SVG-only and SVG\,+\,Image both produce component-mapped selectors, with SVG\,+\,Image additionally confirming visual appearance and achieving higher-confidence grounding. This pattern is consistent across all four visualizations. SVG\,+\,Image correctly identifies and confirms the encoding while image-only produces generic descriptions without DOM anchors.

\begin{table*}[ht]
\centering
\caption{Interaction step comparison across four visualizations and all four modality conditions.}
\label{tab:merged-interaction}
\renewcommand{\arraystretch}{1.4}
\small
\begin{tabularx}{\linewidth}{p{0.09\linewidth} p{0.12\linewidth} X X X X}
\toprule
\textbf{Visualization}
  & \textbf{Interaction}
  & \textbf{SVG Only}
  & \textbf{Image Only}
  & \textbf{SVG\,+\,Image}
  & \textbf{SVG\,+\,Image\,+\,Interaction Source Code} \\
\midrule

PrettiSmart
& Simulation card &
Hover reveals guide lines and outlined box; align to the same
window in the flow view.
\newline\textit{Basis: DOM structure} &
``Click a thumbnail to switch the central flow.''
\newline\textit{Incorrect: hover confused with click} &
Hover reveals guide lines and outlined box.
\newline\textit{Partially verified; click-to-load consequence
unknown} &
Click a card to load that simulation; timeline and state marks
update.
\newline\textit{Selector: \texttt{g.simu rect.hover.h\_rect} —
D3 verified} \\

& Function card hover &
\textit{Not generated} &
``Use F0, F1, F2 to filter function events.''
\newline\textit{Unverified; visual resemblance only} &
\textit{Not generated} &
Hover a function card to highlight its calls across the
timeline.
\newline\textit{Selector: \texttt{g.func} — D3 mouseover
verified} \\

& Workflow guidance &
\textit{Not generated} &
\textit{Not generated} &
\textit{Not generated} &
Pick a simulation $\rightarrow$ inspect function cards
$\rightarrow$ hover to spotlight calls $\rightarrow$ trace
flows $\rightarrow$ confirm state changes.
\newline\textit{Cross-component event chain — D3 verified} \\

\midrule

RuleMatrix
& Rule row hover &
Hover highlights row; cross-highlight inferred from DOM class.
\newline\textit{Basis: DOM structure} &
``Hover a row to highlight it.''
\newline\textit{Correct but no selector anchor} &
Hover highlights row and cross-highlights linked columns.
\newline\textit{Partially verified; filter consequence unknown} &
Hover a rule row to highlight it and filter the linked output
probability column.
\newline\textit{Selector: \texttt{g.rules rect} — D3 mouseover
verified} \\

& Sort interaction &
\textit{Not generated} &
``Click column headers to sort.''
\newline\textit{Visual affordance — unverified} &
\textit{Not generated} &
Click a column header to re-sort rules by that criterion;
row order updates.
\newline\textit{Selector: \texttt{g.header text} — D3 click
verified} \\

\midrule

PonziLens
& Path selection &
Click path to expand detail view; inferred from DOM nesting.
\newline\textit{Basis: DOM structure} &
``Click a path to see details.''
\newline\textit{Visual affordance — partially correct} &
Click path expands detail; related action dots highlighted.
\newline\textit{Partially verified; cross-component update
unknown} &
Click a path to expand the Execution Detail panel; action dots
update to reflect selected path's event sequence.
\newline\textit{Selector: \texttt{g.overview\_G path} — D3
click verified} \\

& Legend filter &
\textit{Not generated} &
``Click legend items to filter.''
\newline\textit{Visual resemblance — unverified} &
\textit{Not generated} &
Click a legend item (READ/WRITE/CALL) to toggle visibility
of corresponding action dots across all modules.
\newline\textit{Selector: \texttt{g.legend g.WRITE} — D3 click
verified} \\

\midrule

Shanghai Index
& Brush zoom &
Brush on overview panel zooms main chart; inferred from DOM
rect overlap.
\newline\textit{Basis: DOM structure — partially correct} &
``Drag the overview bar to zoom.''
\newline\textit{Visual affordance — correct action, no anchor} &
Brush zooms main candlestick view; MA lines update
accordingly.
\newline\textit{Partially verified; MA line update unknown} &
Drag the brush on the overview panel to zoom the main chart;
all MA lines and candlesticks update to the selected range.
\newline\textit{Selector: \texttt{g[cursor="pointer"]} — D3
brush verified} \\

& Tooltip hover &
\textit{Not generated} &
``Hover to see price details.''
\newline\textit{Visual affordance — unverified} &
\textit{Not generated} &
Hover a candlestick to reveal OHLC values and volume in a
tooltip.
\newline\textit{Selector: \texttt{g[clip-path] > path} — D3
mouseover verified} \\

\bottomrule
\end{tabularx}
\end{table*}

\textbf{Key Findings (Table~\ref{tab:merged-interaction}).}
The interaction results are consistent across all four visualizations. SVG-only infers interactions from DOM structure (e.g., hover overlays, nested click targets) but systematically misses action-consequence relationships that are only encoded in event handlers. For example, it correctly identifies the simulation card hover overlay in PrettiSmart but cannot infer that clicking loads a new simulation and updates the timeline. Image-only produces the least reliable output, as it correctly identifies interactive affordances from visual cues (buttons, draggable regions, hover-styled elements) but confuses interaction types (e.g., click vs.\ hover in PrettiSmart; unverified filter behaviors in RuleMatrix and PonziLens) and produces no DOM valid selectors for any interaction step across any visualization. SVG\,+\,Image partially bridges the gap and it confirms which elements are interactive visually but still cannot infer cross-component consequences without source code. Only the full pipeline (SVG\,+\,Image\,+\,Interaction Source Code) produces all interaction steps with verified selectors and cross-component workflow guidance in all four visualizations, replacing convention-based guesswork with grounded action-consequence reasoning derived directly from event handlers.

In summary, each modality provides a distinct and non-redundant capability across all four tested visualizations: SVG structure offers precise visual component decomposition and DOM-valid selectors; bitmap images resolve spatial ambiguities and confirm visual appearance; and interaction source code replaces convention-based guesswork with verified event handler logic. All three inputs are essential for accurate, grounded, and actionable onboarding, and the necessity of each modality is consistent regardless of composite design pattern or domain complexity.}